\documentclass{jfm}
\usepackage{graphicx}
\usepackage{epstopdf, epsfig}
\usepackage{amsmath}
\usepackage{here}

\title{Response of Non-premixed Jet Flames to Blast Waves}

\author{
    Akhil Aravind\aff{1},
    Gautham Vadlamudi\aff{1}
    \and Saptarshi Basu\aff{1,2}
        \corresp{\email{sbasu@iisc.ac.in}}
    }

\affiliation
{\aff{1} Department of Mechanical Engineering, Indian Institute of Science, Bangalore, India.
\aff{2} Interdisciplinary Centre for Energy Research, Indian Institute of Science, Bangalore, India.}

\begin{document}

\maketitle

\begin{abstract}
The work investigates the response dynamics of non-premixed jet flame to blast waves that are incident along the jet axis. In the present study, blast waves, generated using the wire-explosion technique, are forced to sweep across a non-premixed jet flame that is stabilised over a nozzle rim positioned at a distance of 264 mm from the source of blast generation. The work spans a wide range of fuel jet Reynolds numbers ($Re$) and incident blast wave Mach numbers ($M_{s,r}$). The interaction imposes a characteristic flow field over the jet flame marked by a sharp discontinuity followed by a decaying profile and a delayed second spike. The second spike in the flow field profile corresponds to the induced flow that follows the blast front. While the response of the flame to the blast front was minimal, it was found to detach from the nozzle rim and lift off following the interaction with the induced flow. Subsequently, the lifted flame was found to re-attach back at the nozzle or extinguish, contingent on the operating $Re$ and $M_{s,r}$. Alongside flame lift-off, flame tip flickering was aggravated under the influence of the induced flow. A simplified theoretical model extending the vorticity transport equation was developed to estimate the change in flickering timescales and length scales owing to the interaction with the induced flow. The observed experimental trends were further compared against theoretical predictions from the model. 
\end{abstract}


\section{Introduction}\label{sec:intro}

The interaction between flames and finite-amplitude pressure waves, such as shock or blast waves, has attracted considerable research attention due to its critical role in diverse applications, ranging from propulsion systems to fire safety measures. In burners and combustion chambers, finite-amplitude acoustic oscillations at resonant frequencies often develop due to repeated reflections along the chamber boundaries (\cite{dobashi_flame_1994,jones_pressure_1991}). These finite amplitude disturbances tend to distort and deform the flame inside the combustor. Contingent on the operating flow conditions (turbulence intensity, boundary conditions), the domain geometry, and reactivity of the fuel, this can lead to an increase in the local burning rates and can eventually transition the flame (deflagration wave) into a detonation front (\cite{thomas_experimental_2001}). Such transitions are vital in shaping the combustion dynamics of propulsion systems such as pulse detonation engines (\cite{nikitin_pulse_2009}), which rely on detonations for efficient combustion and higher thrust. However, on the other hand, shock-flame interactions can also destabilise the flame and drive it towards extinction (\cite{yoshida_blowoff_2024}). This principle is utilized in large-scale fire extinguishing systems to fight wildfires and oil fires. They employ blast waves generated from explosives to blow off flames from the fuel source and extinguish them (\cite{akhmetov_extinguishing_1980, akhmetov_formation_2001, yoshida_blowoff_2024}). 

Traditionally, fundamental experimental studies on the interaction of flames with non-linear pressure waves were performed inside a shock tube or a combustion chamber filled with a combustible mixture. Typically, the shock/blast front would be initiated at a chamber boundary after a characteristic time delay following the ignition of the mixture at a desired location inside the chamber. Pioneering work of \cite{markstein_shock-tube_1957} investigated the distortion of curved stoichiometric butane-air flames as a planar shock front swept across it from the unburnt side to the burnt side. He observed the emergence of an unburnt gas funnel that swiftly penetrated into the burnt side, inducing a reversal of the flame front and causing a rapid transition into fine turbulent structures. Recent studies from \cite{la_fleche_dynamics_2018} reported similar observations following the interaction of a cellular flame front with a blast wave in a Hele-Shaw cell. The cellular flame was found to exhibit significant distortions and a reversal of the flame front as the blast wave swept through it from the unburnt side to the burnt side. These effects were attributed to the misalignment between the density and pressure gradient fields that contributes to local baroclinic vorticity production.

Further studies by \cite{scarinci1993amplification} showed that the reaction rates were accelerated due to the enhanced vorticity production and that distortion of the flame front at larger time scales was contingent on the reactivity of the mixture. While a highly reactive mixture of acetylene and air was found to re-establish its initial flame shape owing to the increased reaction rates, which in turn caused a rise in viscous dissipation effects due to higher temperatures, a less reactive mixture like methane-air was found to retain its deformed shape for longer timescales. Numerical studies by \cite{batley1994numerical,batley1996baroclinic} traced out the increment in the reaction rates against baroclinic vorticity production in a cylindrical laminar premixed flame following its interaction with a planar pressure wave. \cite{thomas_experimental_2001} extended the above studies and demonstrated that multiple interactions of the flame with shocks can cause a substantial increment in the local temperature and pressure (accompanied by increased reaction rates) and can transition the flame into a detonation front. 

While the aforementioned fundamental studies have significantly advanced our comprehension of flame interactions with nonlinear pressure waves, most studies focus on canonical configurations involving shocks generated in a confined chamber (which can be approximated to near-steady shock or unsteady shocks at low decay rates) and geometrically simple flame structures (planar/cylindrical/spherical flame front). Moving beyond canonical configurations, \cite{chan_interactions_2016} investigated shock-flame interactions in more practical burner settings. Their experiments employed non-premixed flames ranging from small Bunsen burners to large-scale ring burners. These flames were subjected to a transverse high-speed exhaust flow generated by a shock tube, which was characterised by a planar shock front followed by a high-momentum compressible vortex ring. Interestingly, the shock front itself caused minimal disruption to the flame. However, the subsequent induced flow was found to drive the flame to extinction. The study documented the minimal transverse velocity required to blow off the flame.

Building upon these studies, the current work explores the axial interaction between non-premixed jet flames and blast waves. The study employs a miniature blast wave generator designed on the principle of high-voltage wire explosion (\cite{oshima_blast_1962}). The technique involves imposing a high-voltage electrical impulse onto a thin metallic wire, causing it to vaporise instantaneously and form a dense metallic vapour cloud. The expansion of this high-temperature, high-pressure vapour column results in the formation of a blast wave. The initial blast front generated at the wire is cylindrical in nature, owing to the wire acting as a line source for the explosion. However, the blast rapidly transitions into an ellipsoidal form and eventually tends towards spherical symmetry at large propagation distances (\cite{chiu1977blast}). The facility has been extensively used to study the secondary atomisation of liquid droplets at high Weber numbers (\cite{sharma_shock_2021,sharma_shock-induced_2023}). Recent studies by \cite{chandra_shock-induced_2023} validated the experimentally observed evolution of the generated blast wave against the theoretical blast wave model developed by \cite{bach_analytical_1970}. It is to be noted that the generated blast wave is followed by an induced bulk flow, similar to that reported by \cite{chan_interactions_2016}. Thus, the effective velocity/pressure profile exhibits two peaks, one corresponding to the blast front and the other corresponding to the subsequent induced flow. When this flow field is imposed on the jet flame, the large spatiotemporal flow field gradients are expected to drastically alter its response dynamics compared to that observed in canonical shock/blast-flame interaction settings. 

In addition to the expected distortion of the flame boundary following the interaction with the blast wave and the subsequent induced flow, the characteristic flickering instability of non-premixed jet flames is expected to alter under its influence. Flame flickering is a consequence of buoyancy-induced roll-up of toroidal vortices about the shear boundary between the hot product gases and the ambient air. The rolled-up vortices advect downstream and shed (detach from the feeding shear layer) upon reaching a critical circulation limit (\cite{xia_vortex-dynamical_2018}). This process of periodic vortex shedding induces a cyclical stretching of the flame tip, resulting in a visually perceived flickering behaviour. 
Our previous works (\cite{pandey_dynamic_2021, thirumalaikumaran_insight_2022}) investigating the flame dynamics of droplet diffusion flames subjected to external flows have demonstrated that the vortex roll-up rate in the shear boundary is aggravated in the presence of external flows and can even result in flame pinch-off events contingent on the operating conditions. Thus, the blast wave and subsequent induced flow, which impose large spatiotemporal velocity gradients, are expected to drastically alter the dynamics of flame flickering. 

Building upon the arguments presented above, the present study aims to investigate the response dynamics of non-premixed jet flames following their axial interaction with a blast wave and the induced flow subsequent to it. The study classifies the response of the flame into distinct qualitative regimes across the parametric space of fuel jet Reynolds number and blast wave Mach numbers. Extending the theory of flame shedding developed by \cite{xia_vortex-dynamical_2018}, the study proposes a simplified theoretical model to estimate shedding timescales following the interaction process.

\section{Experimental Setup}\label{sec:Expt_setup}

    \subsection{Test Facility} \label{sec:Test_facility}

\begin{figure}
        \centering
        \includegraphics[width=0.9\linewidth]{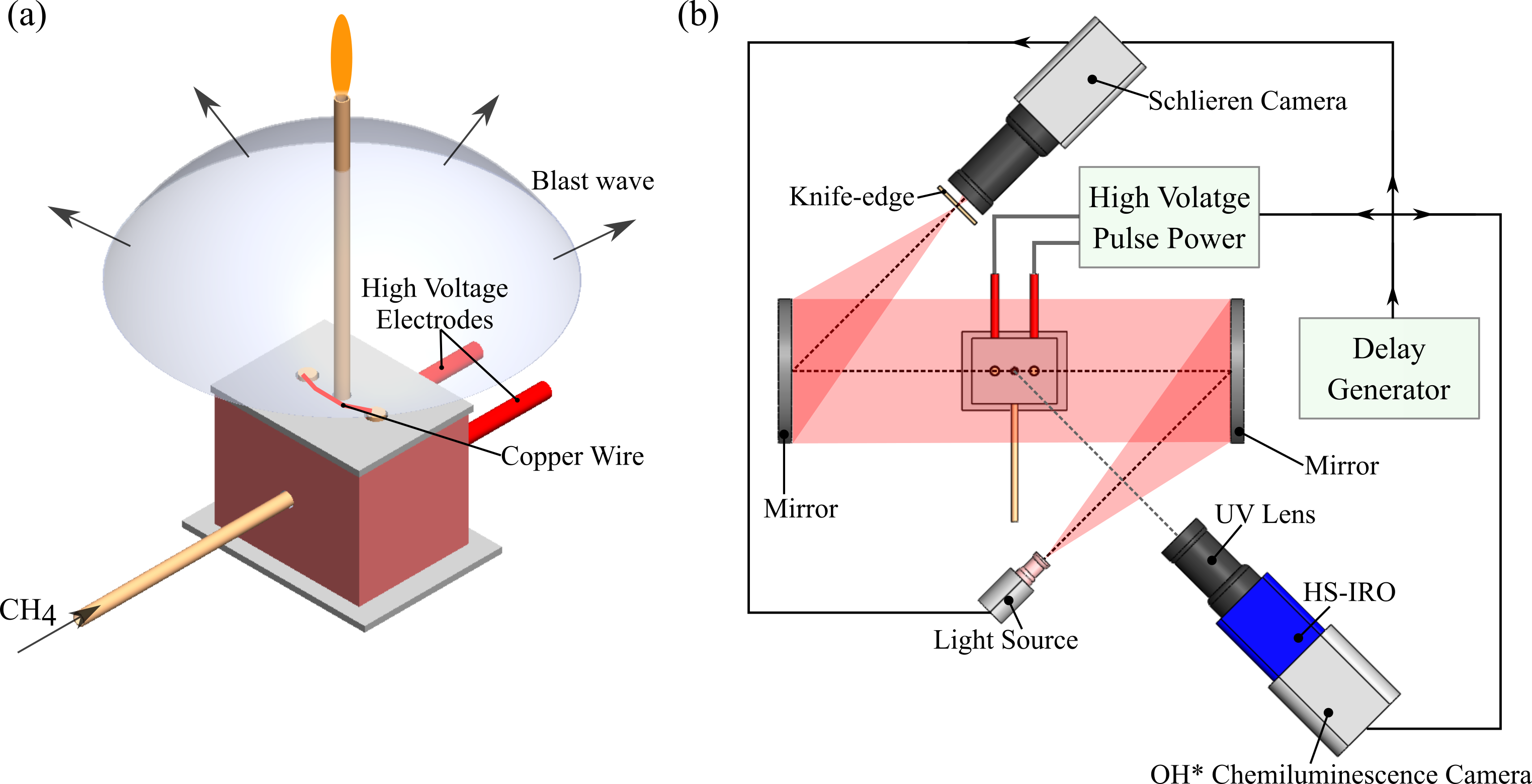}
        \caption{(a) Schematic of the high voltage electrode chamber that is used to generate blast waves. (b) Schematic of the experimental setup and imaging facility used to visualise the interaction between non-premixed jet flames and blast waves.}
        \label{fig:Exp_setup}
    \end{figure}

\begin{figure}
        \centering
        \includegraphics[width=0.9\linewidth]{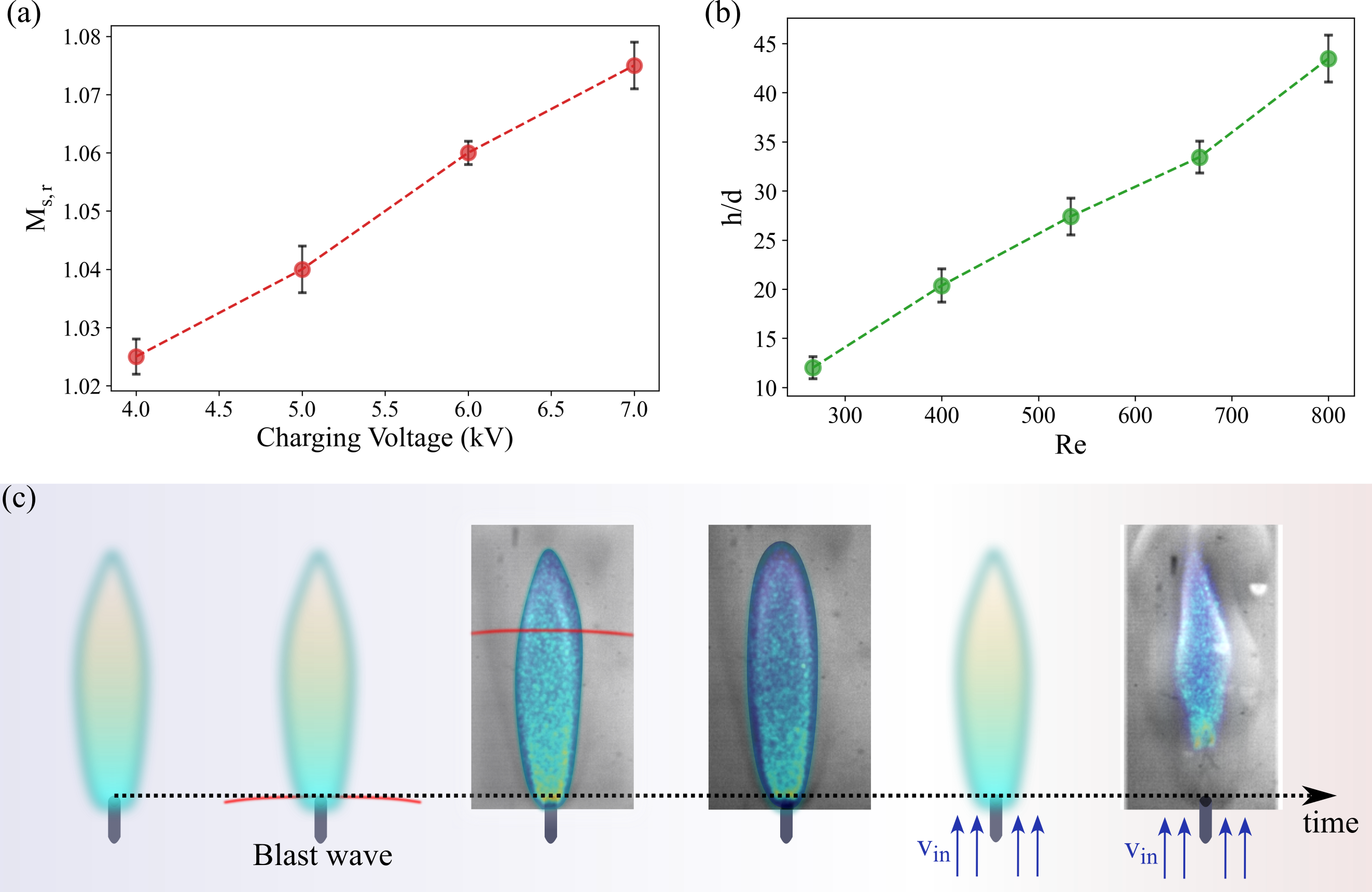}
        \caption{(a) The variation of the reference blast wave Mach number ($M_{s,r}$) at different capacitor charging voltages, (b) The variation of the flame height of the non-premixed jet flame across the parametric space of fuel jet Reynolds numbers.}
        \label{fig:Blast_Flame_Characterisation}
    \end{figure}

    The experimental facility employed in the current work comprises a specially designed blast wave generator and a fuel jet nozzle built into it to facilitate the axial interaction between non-premixed jet flames and blast waves. The system is schematically depicted in  Figure \ref{fig:Exp_setup}(a). The blast wave generator operates on the principle of high-voltage wire explosion and has been extensively used for shock/blast generation over the past few decades, particularly for droplet atomization studies (\cite{sharma_shock_2021,chandra_shock-induced_2023,sharma_shock-induced_2023}). When a high-voltage electrical pulse of the order of kilovolts is imposed across a thin metallic wire, it results in rapid joule’s heating of the wire, causing it to instantaneously melt and vaporise, thereby forming a dense column of the metal vapour. The expansion of this vapour column generates the blast front (\cite{oshima_blast_1962}). In the current system, a 2kJ power pulse generator (Zeonics Systech, India Z/46/12) that houses a 5$\mu$F capacitor is used to provide the high-voltage pulse across two electrodes that are connected through a thin metallic copper wire (35 SWG, Length, $L_{w}$=75mm). The charging voltage of the capacitor is varied between 4kV to 7kV, resulting in blast waves with Mach numbers ($M_{s,r} = {u_{s}}/{c}$) spanning from 1.025 to 1.075. Here, $u_{s}$ is the speed of the blast front as measured at its instant of interaction with the non-premixed flame at the nozzle exit (detailed in Section \ref{sec:Data_Processing}), and $c$ is the local speed of sound under ambient conditions (Temperature of 298K and pressure of 1atm). The plot depicting the dependence of the Mach number of the generated blast front ($M_{s,r}$) on the capacitor charging voltage is provided in Figure \ref{fig:Blast_Flame_Characterisation}(a). The blast wave imposes a characteristic flow field profile on the jet flame marked by a sharp discontinuity, followed by a decaying profile that drops to sub-ambient levels within a specific time scale. Nonetheless, an induced bulk flow trails behind the blast front with a characteristic time delay, introducing an additional flow component that modifies the dynamics of the jet flame (illustrated in \ref{fig:Blast_Flame_Characterisation}(c)). These observations are elaborated upon extensively in sections \ref{sec:Flow_field} and \ref{subsec:global Obs}.

    A fuel jet nozzle tube, with an inner diameter ($d$) of 2 mm, is centrally positioned with respect to the electrode base plate. This base plate forms the section of the electrode chamber over which the electrodes are connected with the thin copper wire. As depicted in  Figure \ref{fig:Exp_setup}(a), the nozzle tube, positioned equidistantly relative to both electrodes, conveys methane gas. A precise mass flow controller (Bronkhorst Flexi Flow Compact, capacity: 0-1.6 SLPM) regulates the methane feed. The nozzle stabilises a non-premixed jet flame at its exit, located 264 mm above the base plate. In the current study, the fuel-jet velocity ($v_{j}$) is varied between 1m/s to 3m/s in steps of 0.5m/s, varying the fuel jet Reynolds number ($Re = {v_{f}d}/{\nu_{f}}$) from 267 to 800, in steps of 133. The flame height (estimation presented in Section \ref{sec:Data_Processing}) was found to increase with an increase in the fuel jet Reynolds numbers, and their dependence is depicted in Figure \ref{fig:Blast_Flame_Characterisation}(b). 
    
    During the experimental runs, the fuel jet is set to a desired Reynolds number, and the non-premixed jet flame is stabilised at the nozzle exit using a pilot flame. Simultaneously, the 5$\mu$F capacitor is charged to a required voltage level contingent on the operating Mach number. Once charged, the charging circuit is cut off. The discharge circuit running through the electrodes and the thin copper wire is triggered upon receiving a TTL pulse from a digital delay generator (BNC 757) that syncs the blast generation process with the imaging systems (see Figure \ref{fig:Exp_setup}(b)). 
    
    The interaction process was investigated using a combination of Schlieren flow visualisation and OH* chemiluminescence imaging. The Schlieren setup utilised a pair of parabolic concave mirrors with a focal length of 1500 mm. A high-speed, non-coherent pulse diode laser (Cavitar Cavilux smart UHS, 400 W), emitting at a wavelength of 640 nm, served as the light source (depicted in Figure \ref{fig:Exp_setup}(b)). Schlieren visualisation was performed at 40000 fps using a High-speed star SA5 Photron Camera. Simultaneous OH* chemiluminescence imaging was performed using the LaVision SA5 High-speed camera, coupling it with a high-speed intensifier (LaVision HS-IRO, IV Generation), a UV lens (Nikon Rayfact PF10445MF) and an OH* bandpass filter ($\sim$ 310 nm). The acquisition rate was set to 10000 fps for OH* chemiluminescence imaging. The spatial resolutions for Schlieren visualisation and OH chemiluminescence imaging were set to 5.012 px/mm and 3.430 px/mm, respectively.

    \subsection{Data Processing} \label{sec:Data_Processing}

    OH* chemiluminescence images of the flame were analysed using the ImageJ software. The Otsu thresholding technique, integrated within ImageJ, was employed to identify the flame boundary. This method effectively segregates the image pixels as foreground (flame) and background based on an intensity threshold ($I_f$). This threshold is determined by minimising the variance within each category (foreground and background), effectively maximising the variance between them. Pixels with intensities exceeding $I_f$ are assigned a binary value of 1 (designated as flame), while those falling below the threshold are assigned a value of 0 (designated as background). The resulting contiguous region of pixels with a value of 1 subsequently demarcates the flame perimeter.
    
    Following blast wave interaction, the spatiotemporal response dynamics of the flame base and flame tip were tracked individually. Flame height was estimated as the vertical distance between the flame tip and the flame base. The OH* chemiluminescence signal, indicative of the flame's overall heat release rate, was obtained by integrating the pixel intensities within the delineated flame boundary in the original image. To ensure reliable data, all flame descriptors presented hereafter (flame tip and base positions, height, and OH* chemiluminescence signal) represent the average of at least three independent experimental trials.

    \begin{figure}
        \centering
        \includegraphics[width=0.9\linewidth]{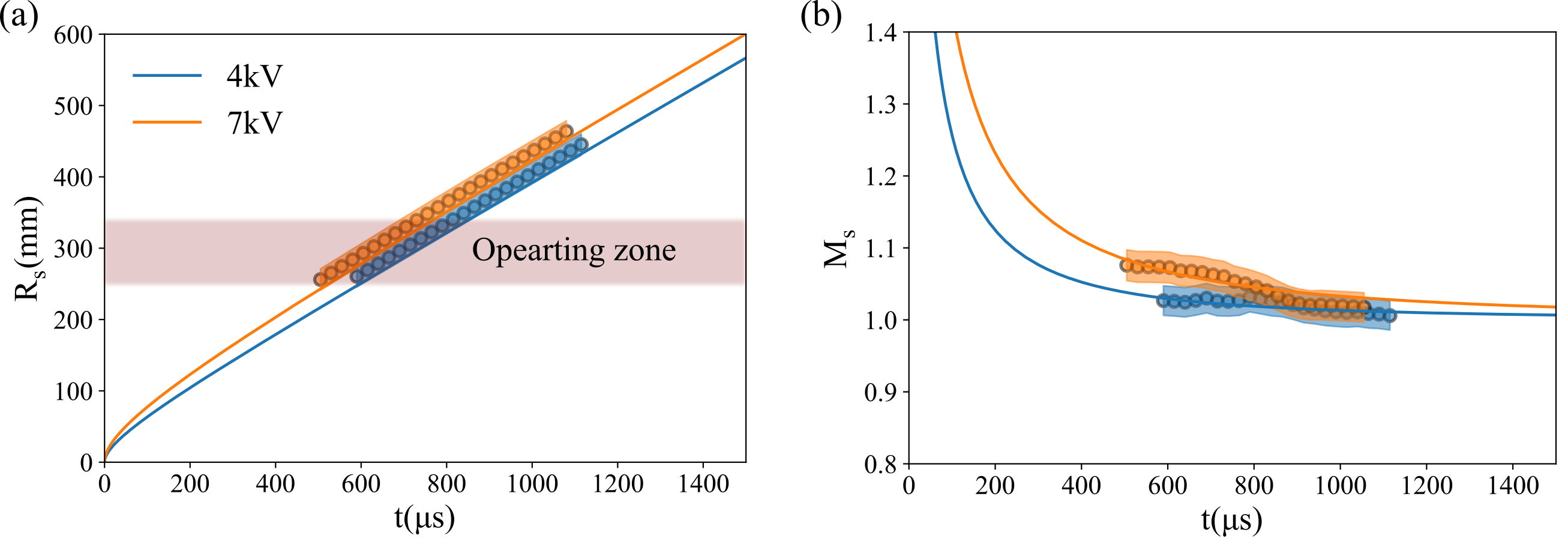}
        \caption{This figure presents a comparison between the experimentally measured temporal evolution of the blast wave at charging voltages of 4 kV and 7 kV against the analytical blast wave model proposed by \cite{bach_analytical_1970}. Panels (a) and (b) depict the trends observed in the blast wave radius and Mach number, respectively.}
        \label{fig:Blast_Validation}
    \end{figure}
    
    Schlieren imaging provided a means to visualise the evolution of the blast wave. By spatially tracking the position of the blast wave (along the axis of the fuel jet) with respect to time in Schlieren images, a time-dependent variation of the blast wave radius ($R_s$) was obtained. The data was then used to estimate the velocity of the blast front ($u_{s}$) and, subsequently, the blast wave Mach numbers ($M_{s}$).

    \subsection{Characterization of the flow-field imposed by the blast wave} \label{sec:Flow_field}

    Owing to the cylindrical geometry of the exploding wire, the initial blast front is expected to exhibit a cylindrical form. Studies suggest that these cylindrical blast fronts initially transition to ellipsoidal forms, ultimately evolving to spherical fronts in the far field. However, at blast propagation distances ($R_{s}$) of the order of the characteristic length of the exploding wire ($L_{w}$), the ellipsoidal nature of the blast front is found to be retained (\cite{chiu1977blast}). Our experimental data on the temporal evolution of the blast wave radius and Mach number is also found to align more favourably with the theoretical model of an expanding cylindrical blast wave (\cite{bach_analytical_1970}) compared to a spherical blast wave (Supplementary Section S1). This consistency suggests that during the blast-flame interaction process, the cylindrical nature of the blast wave is likely preserved in the immediate vicinity of the fuel jet axis. Furthermore, the observed ratio of the blast wave radius and the characteristic wire length ($R_{s}/L_{w}$) of $\sim 3.5$ in our experiments further supports the notion of a cylindrical blast front interacting with the jet flame. Details illustrating the implementation of the theoretical blast wave model for the present experimental configuration are presented in Supplementary Section S1. 

\begin{figure}
        \centering
        \includegraphics[width=0.9\linewidth]{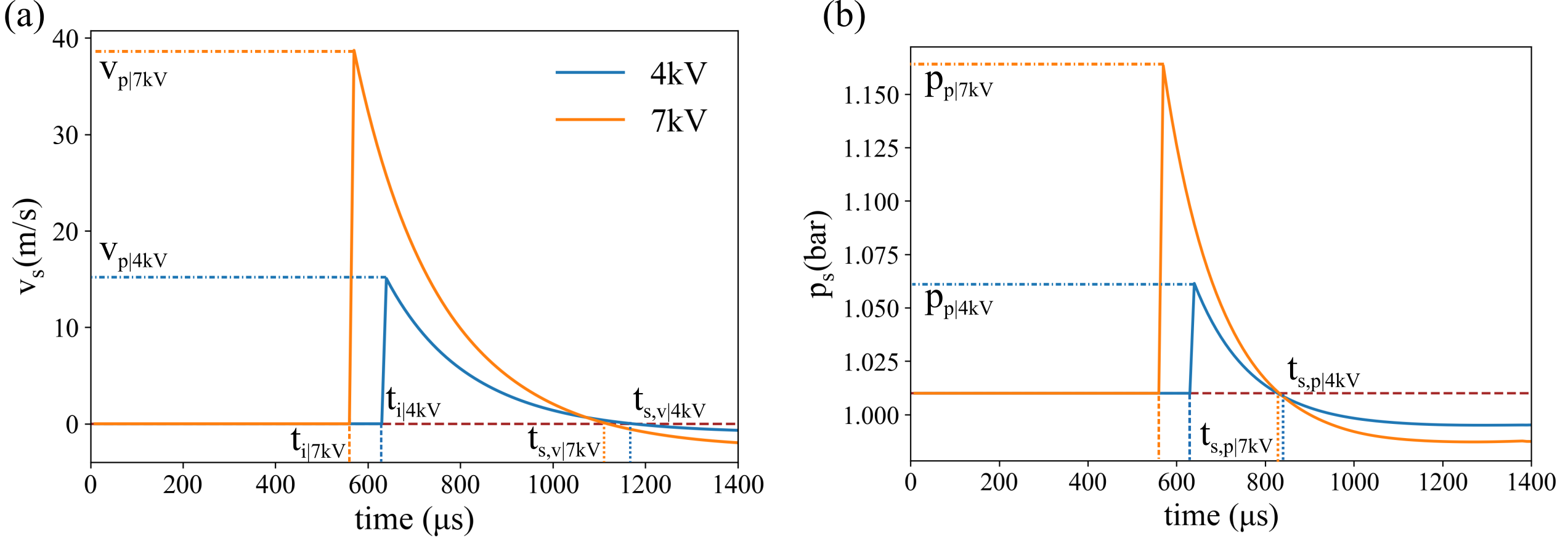}
        \caption{The figure plots the theoretical flow field profiles imposed by the blast wave at the nozzle exit location. The profiles correspond to charging voltages of 4kV and 7kV. Panels (a) and (b) depict the blast-imposed velocity and pressure fields, respectively.}
        \label{fig:Vel_Pre_Profiles}
    \end{figure}

    It is to be noted that the theoretical blast model (\cite{bach_analytical_1970}) relies on an assumed power profile for the density field behind the blast front and estimates the velocity and pressure fields using conservation equations. Additionally, the model assumes that the total mass and energy contained within the blast front remain constant during the evolution process, effectively neglecting entrainment from the surroundings, which is anticipated when the blast-imposed static pressure field drops below ambient levels. The theoretical estimates of the velocity and pressure fields imposed by a cylindrical blast wave at the nozzle exit (radial distance of 264mm from the source of the explosion) are plotted in Figure \ref{fig:Vel_Pre_Profiles}(a,b), respectively. The blast wave is found to reach the nozzle exit location after a time of $t_{i}$ from the time of the explosion. Thus, the blast wave interacts with the non-premixed jet flame (that is stabilised at the nozzle exit) beyond $t_{i}$.  The blast-imposed pressure field attains a peak ($p_{p}$) at the instant of interaction ($t_{i}$) and is followed by a decaying profile that attains sub-ambient levels beyond the time of $t_{s,p}$ (Figure \ref{fig:Vel_Pre_Profiles}(b)). A similar trend is observed for the blast-induced velocity field (Figure \ref{fig:Vel_Pre_Profiles}(a)), wherein the velocity reaches a peak value ($v_p$) at $t_{i}$ and then decays to sub-ambient levels at $t_{s,v}$. It's important to note that beyond $t_{s,p}$, entrainment from the surrounding medium is anticipated, and this renders the analytical solution inaccurate in predicting the flow field. 

    Alongside the blast wave, the flame responds to the delayed, induced bulk flow that follows it. Induced flow following the blast front is a characteristic feature of expanding shock/blast waves and has been reported earlier as ‘blast wind’ by \cite{chan_interactions_2016}. Due to the limitations in experimentally measuring the detailed velocity profile of the induced flow, we could only estimate its characteristic velocity scale based on the flame's response following the interaction with it (detailed in the following section).

\section{Results and Discussions}\label{sec:results}

    \subsection{Global Observations}\label{subsec:global Obs} \addvspace{10pt}

\begin{figure}
        \centering
        \includegraphics[width=0.8\linewidth]{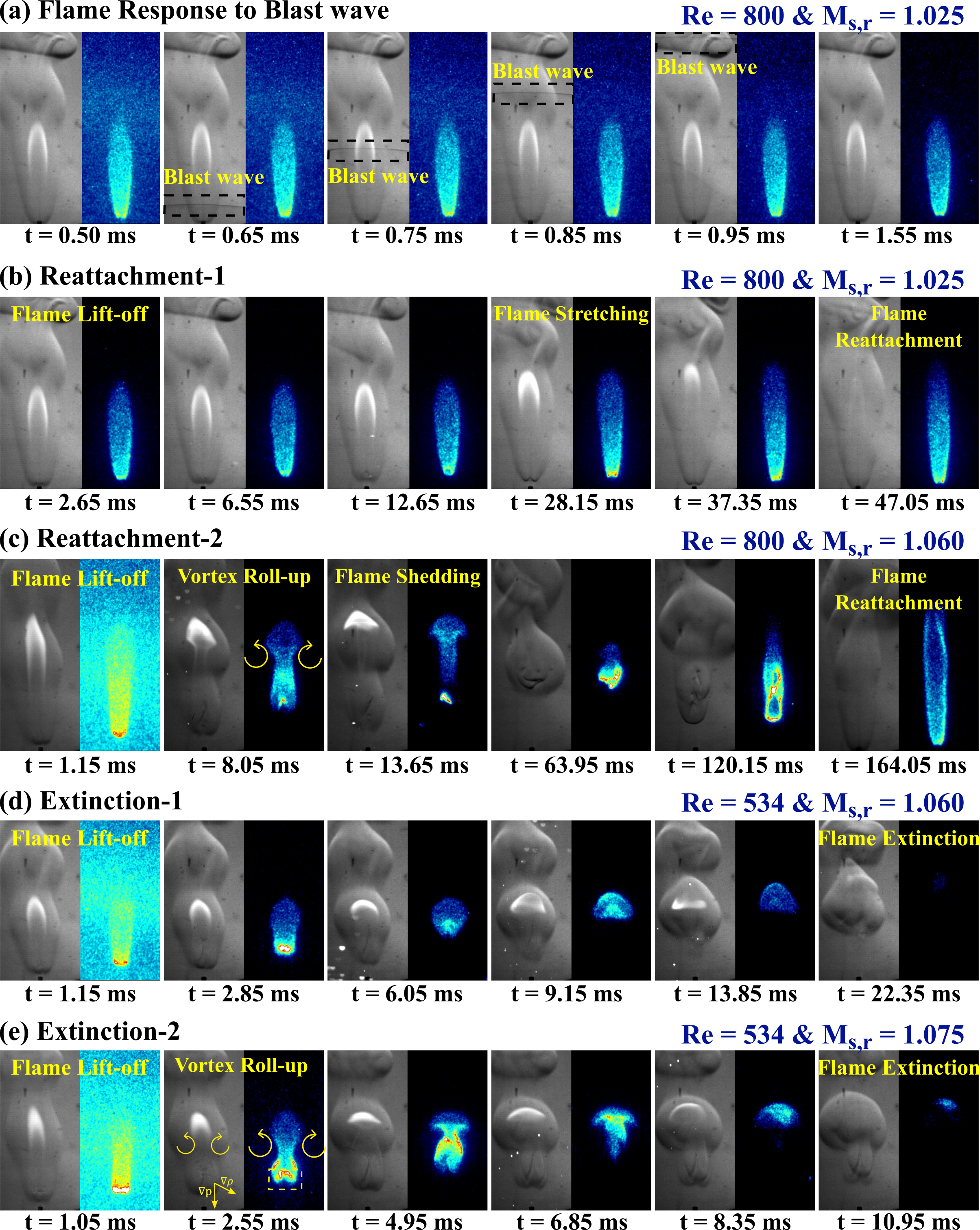}
        \caption{This figure depicts the response of a non-premixed jet flame to blast waves and the subsequent induced flow, categorized by two distinct time scales: the initial flame response at times lower than $t_{s,p}$ (panel (a)), and the response to the induced flow with sub-regimes of reattachment (Type-1 and Type-2, panels (b) and (c)) and extinction (Type-1 and Type-2, panels (d) and (e)) at times greater than $t_{s,p}$. Supplementary movies (1-4) correspond to the depicted sub-regimes in panels (b-e), respectively.}
        \label{fig:Global_All_Flames}
    \end{figure}

    As the blast wave traverses through the non-premixed jet flame, it imposes a characteristic flow field on the jet flame, described by a sharp peak followed by a decaying phase, as outlined in the previous section (Figure \ref{fig:Vel_Pre_Profiles}(a); $t_{i}<t<t_{s,p}$). Additionally, the induced flow interacts with the jet flame after a characteristic time delay following the passage of the blast front. Figure \ref{fig:Global_All_Flames} illustrates the flame's response to the blast front and the subsequent induced bulk flow. The response at timescales lower than $t_{s,p}$ is presented in Figure \ref{fig:Global_All_Flames}(a), while Figure \ref{fig:Global_All_Flames}(b-e) depict the response at timescales beyond $t_{s,p}$. 

    Observations reveal that the flame exhibits minimal response to the initial blast front and the decaying flow field behind it (Figure \ref{fig:Global_All_Flames}(a)). At timescales below $t_{s,p}$, the flame displays only a jittery motion. However, the flame base is found to lift off from the nozzle exit at $t = t_{0}$, where $t_{0}>t_{s,p}$ (Figure \ref{fig:Global_All_Flames}(b-e); first frame). This response corresponds to the induced flow following the blast wave. It is interesting to note that the flame base lift-off rate, at the instant of lift-off, $t_{0}$, remains a near-constant value across the investigated range of fuel jet Reynolds numbers (Figure \ref{fig:Induced velocity schematic}(b)). The plot also reveals a positive correlation between the observed flame base lift-off rate at $t_{0}$ and the strength of the incident blast wave. Anticipating a higher velocity scale for the induced flow at higher Mach numbers, the observed trend in {Figure \ref{fig:Induced velocity schematic}(b) can be attributed to the increased induced flow velocities at higher Mach numbers. 
    
    \begin{figure}
        \centering
        \includegraphics[width=0.9\linewidth]{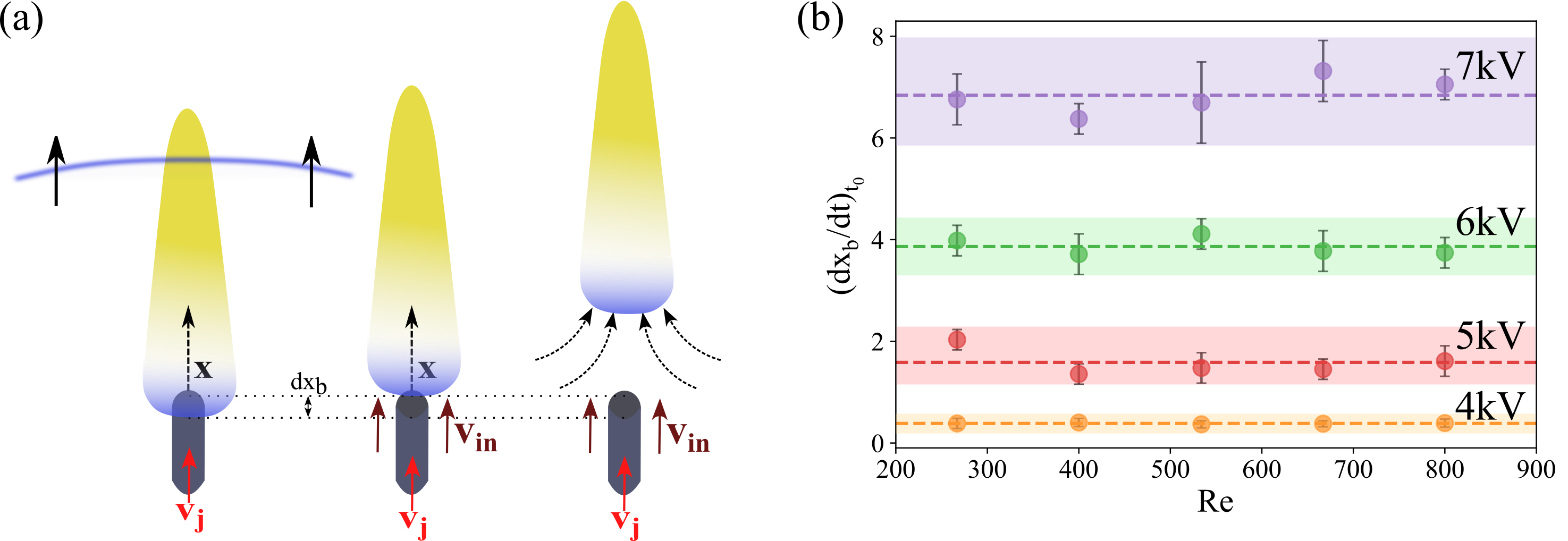}
        \caption{(a) Schematic depicting the lift-off of the jet flame following the interaction with the induced flow. (b) Plot illustrating the flame base lift-off rate at the instant of interaction with the induced flow ($t_{0}$) across the parametric space of fuel jet Reynolds number.}
        \label{fig:Induced velocity schematic}
    \end{figure}
    
    Therefore, we propose the following hypothesis: before $t_{0}$, the non-premixed jet flame remained stabilised on the nozzle rim within a narrow mixing layer between the fuel jet and the surrounding ambient air, wherein heat losses locally balance out the energy generated at the flame front (\cite{li_stabilization_2022}). Upon encountering the induced flow, this local energy balance within the mixing layer is disrupted, and the flame's stoichiometric plane tends to convect with the induced flow (Figure \ref{fig:Induced velocity schematic}(a)). Thus, the estimated flame base lift-off rate at the instant of lift-off ($t_{0}$) can be approximated as the velocity scale of the induced flow that interacts with it. The variation of these induced flow velocity scales across the parametric space of $M_{s,r}$ is plotted in Figure \ref{fig:Induced_Flow_Var_Profile}(a).

    \begin{gather*}
        v_{in} \sim \Bigl( \frac{dx_{b}}{dt} \Bigr)_{t_{0}}
    \end{gather*}
    
    Following the flame base lift-off, the flame is found to exhibit two global response behaviours: re-attachment and extinction. In the re-attachment regimes (Figure \ref{fig:Global_All_Flames}(b,c)), the flame base is found to attain a maximum lift-off height of $h_{b,lft}$ in a period of $t_{b,lft}$, and then re-attach back at the nozzle rim in a characteristic time period of $t_{b,ra}$. However, contingent on the operating fuel jet Reynolds number and the strength of the incident blast wave, the flame tip response dynamics vary. In the reattachment-1 (Type-1) regime, the flame tip exhibits mild distortion and stretching owing to the interaction with the induced flow (Figure \ref{fig:Global_All_Flames}(b)). However, in the reattachment-2 (Type-2) regime (Figure \ref{fig:Global_All_Flames}(c)), the flame tip is found to exhibit a pinch-off event following neck formation at the flame tip. Further details and trends observed in the response dynamics of the flame base and flame tip in the reattachment regimes are detailed in section \ref{subsec:Reattachment}. It is to be noted that as the flame base lifts off following the interaction with the induced flow, ambient air is entrained into the fuel-jet stream (Figure \ref{fig:Induced velocity schematic}(a)). As the fuel and co-axial air streams mix into each other, the lifted flame is expected to attain an edge flame configuration at the flame base (\cite{buckmaster_edge-flames_2002,vadlamudi_insights_2023}). Since the lifted flame tends to reattach back at the nozzle rim in a time period of $t_{b,ra}$, it can be hypothesised that at time scales of the order of $t_{b,lft}$ (at which the flame base lift-off height is maxima), the induced flow field decays to ambient levels and the flame reattaches back at the nozzle rim, following the upstream propagation of the edge flame. Thus, based on the estimated timescale ($\sim O(t_{b,lft})$) and velocity scale ($\sim ({dx_{b}}/{dt})_{t_{0}}$) of the induced flow, a schematic of the anticipated induced flow profile is sketched in Figure \ref{fig:Induced_Flow_Var_Profile}(b).  It is to be noted that the decay timescale of the induced flow ($\sim O(t_{b,lft})$) is at least an order higher in comparison with the decay timescales associated with the blast wave ($t_{s,v}$ and $t_{s,p}$). 

    \begin{figure}
        \centering
        \includegraphics[width=0.9\linewidth]{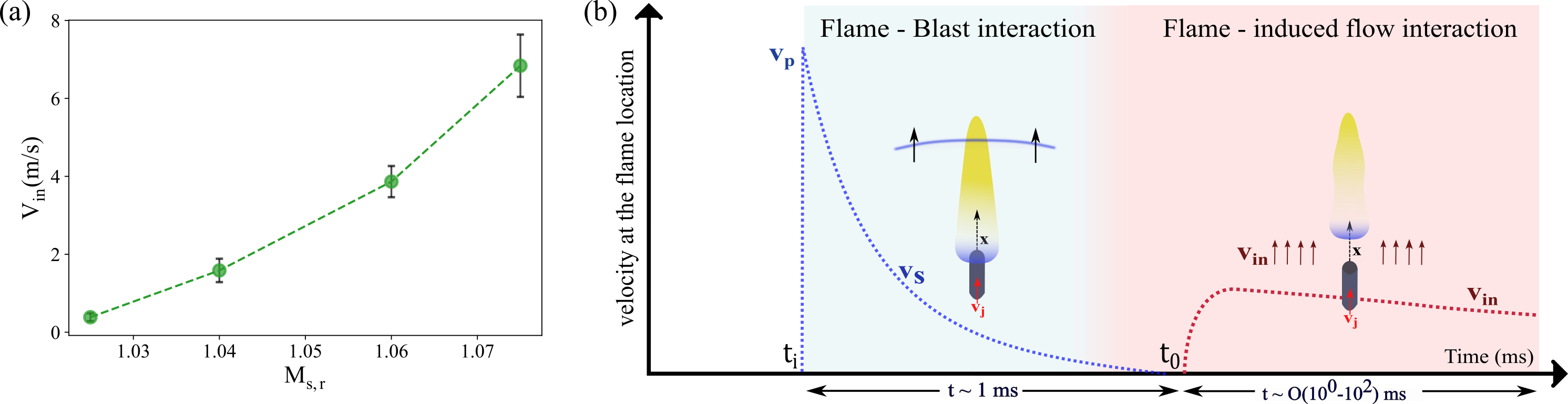}
        \caption{(a) Variation of the estimated induced flow velocity scale ($v_{in}$) across $M_{s,r}$. (b) Schematic depicting the velocity profile at the nozzle exit location imposed by the blast wave and the induced flow following it.}
        \label{fig:Induced_Flow_Var_Profile}
    \end{figure}

    Another global flame response behaviour that was observed was flame extinction. In this regime, the flame base, which lifts off following the interaction with the induced flow, would continue its lift-off without reaching a maximum. This results in an eventual extinction, as illustrated in Figure \ref{fig:Global_All_Flames}(d,e). Similar to the reattachment regimes, Type-1 and Type-2 response behaviours were also noted in the extinction regimes. In the Type-1 sub-regime, flame tip distortions are minimal (see Figure \ref{fig:Global_All_Flames}(d)), while in the Type-2 sub-regime, a notable flame necking event occurs (see Figure \ref{fig:Global_All_Flames}(e)). However, unlike the Reattachment-2 regime, flame pinch-off was not consistently observed throughout the Extinction-2 regime. These trends are further discussed in Section \ref{subsec:Extinction}.

    As the blast wave and the induced flow subsequent to it traverse through the non-premixed jet flame, it imposes a pressure gradient at the flame location oriented upstream along the jet axis. However, the density boundary at the interface of the hot product gases and the ambient air has a component directly radially outward, orthogonal to the fuel jet axis. This results in a baroclinic torque that causes the flame front to exhibit a reversal in the curvature of the flame boundary (see Figure \ref{fig:Global_All_Flames}(e); t = 2.55 ms). These effects are found to be dominant at higher blast strengths as they impose a stronger pressure on the density interface. The observation is similar to that reported by \cite{markstein_shock-tube_1957} and \cite{la_fleche_dynamics_2018}, wherein the concavity of the flame boundary was reversed as a shock wave passed across it from the unburnt side to the burnt side. The effects are attributed to the Richtmyer–Meshkov instability and Rayleigh-Taylor instability along the density boundary surrounding the flame.

    \subsection{Reattachment Response Regimes}\label{subsec:Reattachment} \addvspace{10pt}

    Figure \ref{fig:R1_R2_Quant}(a,b) depicts the flame response in the re-attachment sub-regimes, Type-1 and Type-2, respectively. The plots in the figure track the temporal evolution of the flame base and flame tip alongside the OH* Chemiluminescence signal of the flame. As explained earlier in section \ref{subsec:global Obs}, the flame base tends to lift off from the nozzle rim following the interaction with the induced flow. 
    
    \begin{figure}
        \centering
        \includegraphics[width=0.9\linewidth]{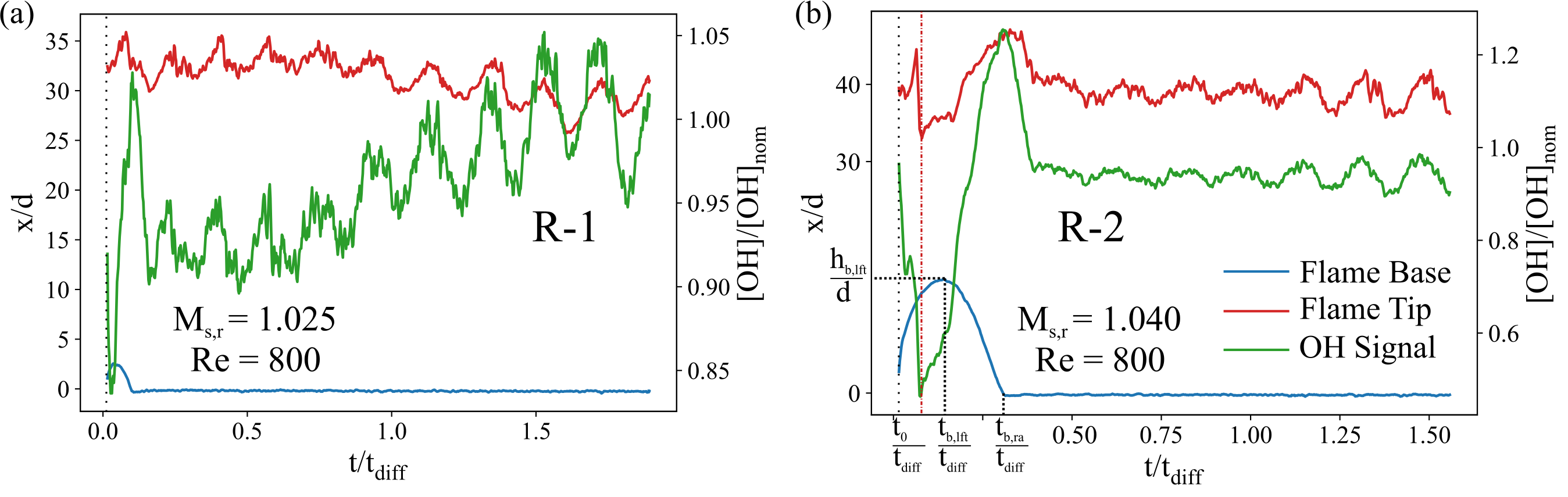}
        \caption{The figure plots the temporal response of the flame base (blue) and the flame tip (red) alongside the OH* chemiluminescence signal of the jet flame as it interacts with the blast wave and the subsequent induced flow. Panels (a,b) correspond to the reattachment sub-regimes, Type-1 and Type-2, respectively. In the plots, the spatial position ($x$) is normalised with the nozzle diameter ($d$) and time is normalised with the timescale associated with the diffusion of methane into ambient air over a characteristic length scale of $d$ ($t_{diff}$). Additionally, the OH* chemiluminescence signal of the flame is normalised by its average value observed under nominal conditions.}
        \label{fig:R1_R2_Quant}
    \end{figure}
    
    The lifted flame attains a maximum lift-off height of $h_{b,lft}$ in a time period of $t_{b,lft}$ (Figure \ref{fig:R1_R2_Quant}(a,b)), both of which are found to increase with an increase in the fuel jet Reynolds number ($Re$) and the induced flow velocity ($v_{in}$), as illustrated in Figure \ref{fig:Liftoff_height_time_Reattach}(a,b), respectively. In the plots, $h_{b,lft}$ and $t_{b,lft}$ are non-dimensionalized with $d$ (diameter of the nozzle tube) and $t_{diff}$, respectively. $t_{diff}$ is the diffusion time scale associated with methane (fuel) diffusing into the ambient air over a length scale of $d$ ($t_{diff} = d^{2}/D$, where D is the molecular diffusion coefficient (\cite{langenberg_technical_2020})).

    \begin{figure}
        \centering
        \includegraphics[width=0.9\linewidth]{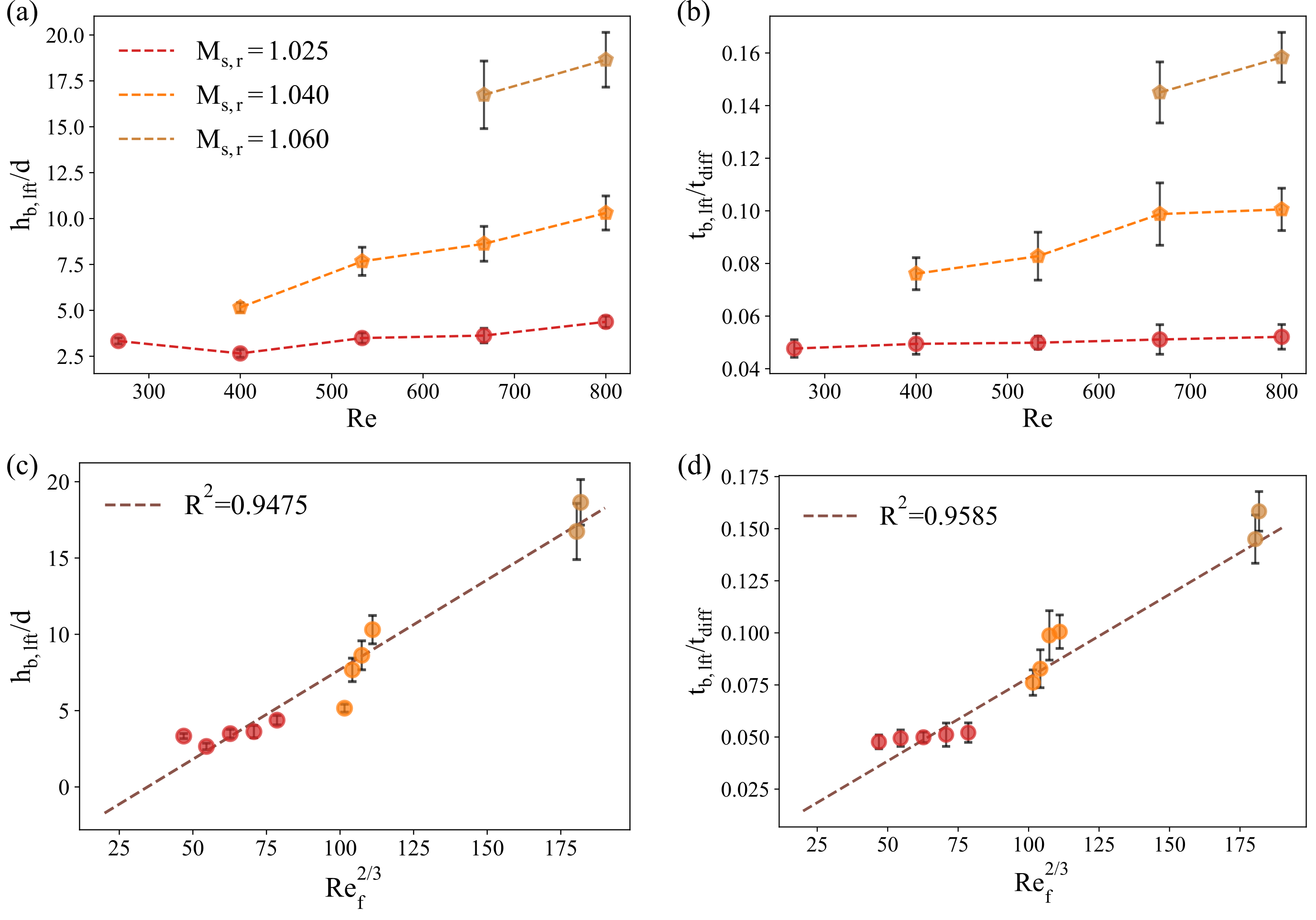}
        \caption{(a,b) The plots depict the variation of the maximum flame base lift-off height ($h_{b,lft}$) and its corresponding timescale ($t_{b,lft}$) across different fuel jet Reynolds numbers, respectively. (c,d) The plots illustrate a linear correlation between $h_{b,lft}$ and $t_{b,lft}$ against $Re_{f}$, respectively. $Re_{f}$ is estimated as an effective velocity scale that accounts for the effects of $v_{f}$ and $v_{ind}$.}
        \label{fig:Liftoff_height_time_Reattach}
    \end{figure}
 
    To better understand the flame base response, we can simplify the interaction between the flame and the induced flow by modelling it as an interaction between the non-premixed jet flame and an impulsively started steady co-axial air flow around the flame, with a velocity scale of $v_{in}$. The modelling is based on the hypothesis that only a small fraction of the induced flow in close proximity to the jet flame interacts with it. This simplification allows us to estimate the effective jet velocity, $v_{f}$, that accounts for the combined momentum of the fuel jet and the co-axial induced flow around it (\cite{alexander_schumaker_mixing_2012}). The flame base lift-off rate of the lifted flame (for $t > t_{0}$) can then be estimated based on $v_{f}$ and the effective upstream flame propagation speed ($S_{L,b}$). It is to be noted that the lifted flame entrains air from the surroundings, develops an edge flame structure at its base, and tends to propagate upstream with an effective flame speed of $S_{L,b}$ while responding to $v_{f}$.

    \begin{gather*}
        \Bigl( \frac{dx_{b}}{dt} \Bigr)_{t>t_{0}} \sim \left[ v_{f}(x,t) - S_{L,b}(x,t) \right] \tag{1}
    \end{gather*}
    
    Our previous work, exploring the response dynamics of premixed jet flames following blast wave interaction (\cite{aravind_response_2024}), has illustrated that the above-mentioned formulation leads to the following correlations:

    \begin{gather*}
        \frac{h_{b,lft}}{d} \sim Re_f^{2/3} \\
        \frac{t_{b,lft}}{t_{diff}} \sim Re_f^{2/3} \tag{2}
    \end{gather*}

    In the formulation, the scales for $h_{b,lft}$ and $t_{b,lft}$ were obtained by estimating $v_{f}(x,t)$ and $S_{L,b}(x,t)$ using appropriate models and imposing the condition that the flame base lift-off rate becomes zero when the flame base attains its maximum lift-off height ($h_{b,lft}$) (see Figure \ref{fig:R1_R2_Quant}(a,b), $t=t_{b,lft}$). Although the formulation was established for premixed jet flames, the correlation was found to be independent of $\phi$ (equivalence ratio) and is expected to hold true for non-premixed jet flames as well. In the scaling law, $Re_{f}$ is the Reynolds number estimated based on the effective jet velocity ($v_{f}$). It is to be noted that for the estimation of $v_{f}$ in the co-axial jet approximation, an effective diameter of the outer co-axial induced flow is calculated, assuming that only a portion of the induced flow in close vicinity to the nozzle tube interacts with the flame. The effective diameter of the co-axial tube conveying the induced flow is approximated to be of the order of the flame base diameter.  

    As illustrated in Figure \ref{fig:Liftoff_height_time_Reattach}(c,d), the above-mentioned correlation holds true for non-premixed flames. The plots illustrate that, at higher effective jet velocities ($v_{f}$), the jet flame is swept to greater downstream distances from the nozzle rim for a longer period of time. 

    Figure \ref{fig:R1_R2_Quant}(a,b) shows that the OH* chemiluminescence signal of the flame exhibits a dip and a subsequent peak following the interaction with the induced flow. This dip corresponds to the phenomenon of flame shedding in jet flames. Flame shedding is associated with the shedding of toroidal vortices that are formed due to gravity-induced shearing at the interface between the hot product gases and the ambient and is observed even under nominal conditions (in the absence of the induced flow).

    To understand the phenomenon of flame shedding better, let us formulate the vorticity transport equation at the shear boundary between the hot product gases and the ambient air (Figure \ref{fig:Shedding_schematic}(a)) for a jet flame established in quiescent conditions.

    \begin{gather*}
        \frac{D\omega}{Dt}=\left(\omega.\nabla\right)u-\ \omega\left(\nabla.u\right)+\ \frac{1}{\rho^2}\left(\nabla\rho\ \text{x}\ \nabla p\right)+\frac{\rho_a}{\rho^2}\left(\nabla\rho\ \text{x}\ g\right)+\ \nu\nabla^2\omega \tag{3}
    \end{gather*}

    In the above equation, $\omega$ represents vorticity, $\rho$ denotes density, $p$ signifies pressure, $u$ is the velocity, $\nu$ is the kinematic viscosity, and $\rho_{a}$ is the density of ambient air. Since the flow is incompressible, axisymmetric and devoid of swirl, the first and second terms on the right side of the equation are zero. As the jet flame is expanding into ambient air, $\nabla p$ is also zero. The diffusion term (the fifth term on the right) describes the re-distribution of vorticity and is not a source of vorticity production (\cite{xia_vortex-dynamical_2018}). However, the fourth term on the right, ${\rho_a}/{\rho^2}\left(\nabla\rho\ \text{x}\ g\right)$, will be non-zero if there is a mismatch in the direction of $\nabla\rho$ with $g$ and can act as a source of vorticity production.

    \begin{figure}
        \centering
        \includegraphics[width=0.9\linewidth]{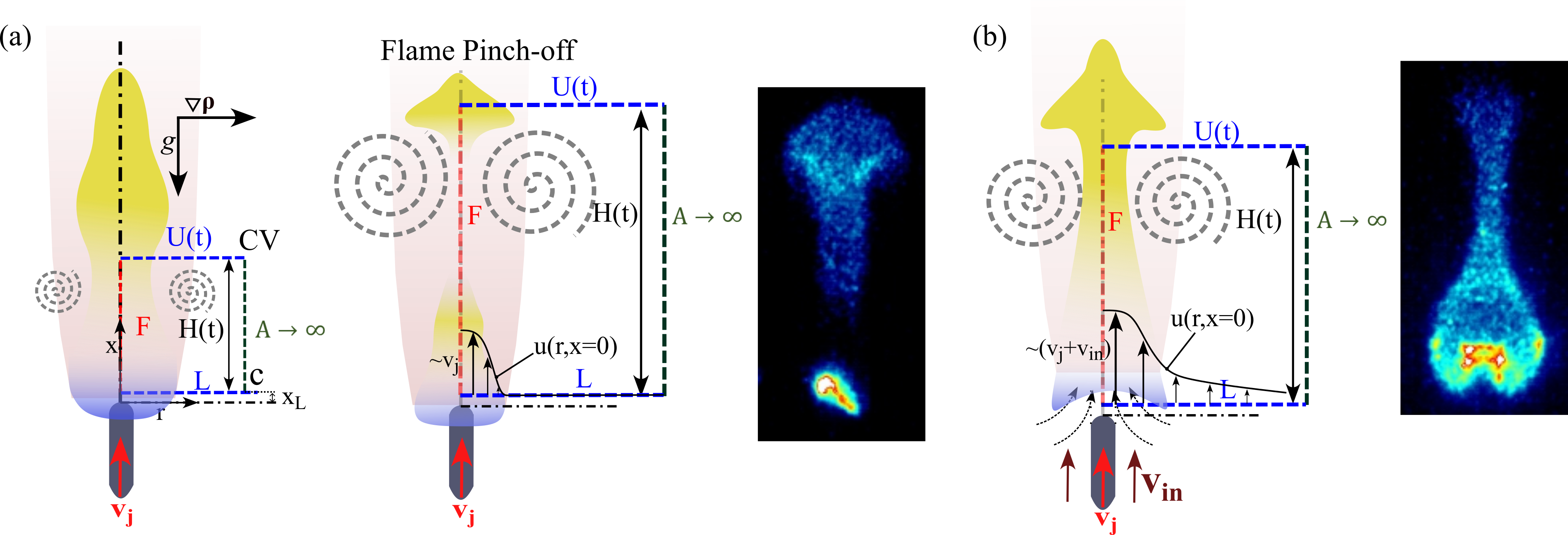}
        \caption{(a) Schematic depicting the shear layer circulation build-up and flame pinch-off in a non-premixed jet flame. Panel (b) depicts the influence of the induced flow on the phenomenon of flame shedding. In the figure, $CV$ is a control volume with a fixed lower boundary ($L$) positioned at an axial distance of $x_{L}$ and a moving upper boundary ($U(t)$) that encloses the toroidal vortex that grows along the shear layer. The right boundary of the control volume ($A$) extends towards the surrounding ambient air, and the left boundary ($F$) is set along the jet axis.}
        \label{fig:Shedding_schematic}
    \end{figure}

    In the present scenario, the body force vector (gravity) acts along the jet axis in the upstream direction (opposite to the fuel jet flow). The density gradient, however, is directed radially outward perpendicular to the fuel jet axis, as illustrated in Figure \ref{fig:Shedding_schematic}(a). Thus, the gradient vectors of the body force field and density field are perpendicular to each other, and this promotes the growth of the toroidal vortices about the shear layer between the hot product gases and the ambient air. The vortices eventually shed upon attaining a critical circulation limit. The shedding of these vortices occurs periodically and is responsible for the periodic flickering pattern at the flame tip (\cite{cetegen_experiments_1993,xia_vortex-dynamical_2018}). Under the influence of the induced flow, the circulation build-up rate of these toroidal vortices is enhanced, and they tend to reach critical circulation limits at shorter time scales. Consequently, this leads to a flame pinch-off event, as explained later in this section. 

    Following the formulation of \cite{xia_vortex-dynamical_2018}, the critical circulation limit at which the toroidal vortices detach from the shear layer and shed can be estimated by evaluating the circulation build-up in a moving control volume enclosing the vortex over a time period of $t_{sh}$ that is associated with vortex shedding. As illustrated in the Figure \ref{fig:Shedding_schematic}(a,b), the control volume, $CV$, evolves in conjunction with the downstream advection of the toroidal vortex. The lower boundary of the control volume, $L$, has a fixed spatial location, while the upper boundary, $U(t)$, moves downstream along with the toroidal vortex it encloses. The height of the control volume is denoted by $H(t)$. 
    Defining $\gamma (x,t)$ as $d\Gamma / dx$, the circulation in the control volume ($\Gamma_{CV}$) can be estimated as         
    \begin{gather*}
        \Gamma_{CV}=\int_{L}^{U(t)} \gamma (x,t) dx \tag{4}
    \end{gather*}    
    The rate of change of circulation within the control volume (simplified using the Leibniz integral rule) can written down as,
    \begin{gather*}
        \frac{d\Gamma_{CV}}{dt} = \int_{L}^{U(t)} \frac{\partial \gamma (x,t)}{\partial t} dx + \gamma (U(t),t) \frac{d U(t)}{dt} \tag{5}
    \end{gather*}
    It is to be noted that the term, $\gamma (L,t) {dL}/{dt}$, from the Leibniz simplification of the integral in Equation (4) is zero since ${dL}/{dt}=0$ (the lower boundary, $L$, of the control volume is fixed in space). A similar formulation can be used to estimate the rate of change of circulation within a control mass, $CM$, that instantaneously coincides with the control volume, $CV$. 
    \begin{gather*}
        \frac{d\Gamma_{CM}}{dt} = \int_{L(t)}^{U(t)} \frac{\partial \gamma (x,t)}{\partial t} dx + \gamma (U(t),t) \frac{d U(t)}{dt} - \gamma (L(t),t) \frac{dL(t)}{dt} \tag{6}
    \end{gather*}
    It is to be noted that ${dL(t)}/{dt}$ for the control mass is not zero since $L(t)$ moves along with $CM$. Equations (5) and (6) can now be used to estimate ${d\Gamma_{CV}}/{dt}$ as, 
    \begin{gather*}
        \frac{d\Gamma_{CV}}{dt} = \frac{d\Gamma_{CM}}{dt} + \gamma (L(t),t) \frac{dL(t)}{dt} \\
        \implies \frac{d\Gamma_{CV}}{dt} = \frac{d\Gamma_{CM}}{dt} + \Bigl(\frac{d\Gamma}{dt}\Bigr)_{L(t)} \tag{7}
    \end{gather*}
    Following the formulation of \cite{didden_formation_1979}, $({d\Gamma}/{dt})_{L(t)}$ can be written down as,
    \begin{gather*}
        \Bigl(\frac{d\Gamma}{dt}\Bigr)_{L(t)} = \int_{L(t)} \omega (x=L,r) u(x=0,r) dr \tag{8}
    \end{gather*}
    The above integral can be simplified as:
    \begin{gather*}
        \Bigl(\frac{d\Gamma}{dt}\Bigr)_{L(t)} = \int_{F}^{A} \omega (x=L,r) u(x=L,r) dr = \int_{F}^{A} u(x=L,r) \left( \frac{\partial u}{\partial r} \right)_{x=L,r} dr \tag{9}
    \end{gather*}
    In the above equation, $\omega = [(\partial u/\partial r) - (\partial v/\partial x)]$, where $u$ and $v$ are the components of the velocities along the axial and radial directions, respectively. $\omega$ in the above equation was simplified, neglecting the effects of $(\partial v/\partial x)$, similar to that employed by \cite{didden_formation_1979}. 
    
    To evaluate the above integral, obtaining the velocity profile along the boundary $L$ is necessary. The velocity profile of an open jet is well-established (\cite{schlichting_laminare_1933}) and is schematically depicted in Figure \ref{fig:Shedding_schematic}(a). The local velocity peaks at $r=0$ and decays to quiescent conditions as $r\rightarrow \infty$. Setting the boundary, $A$, of the control volume at $r \rightarrow \infty$, reduces $({d\Gamma}/{dt})_{L(t)}$ to,
    \begin{gather*}
        \Bigl(\frac{d\Gamma}{dt}\Bigr)_{L(t)} = \int_{0}^{\infty} \frac{\partial}{\partial r} {\left(\frac{u^2}{2}\right)} dr = -\left(\frac{u^2}{2}\right)_{x=L,r=0} \tag{10}
    \end{gather*}    
    Furthermore, ${d\Gamma_{CM}}/{dt}$ in Equation (7) can be simplified as follows,
    \begin{gather*}
        \frac{d\Gamma_{CM}}{dt} = \frac{d}{dt} \int_{c} \Bar{u}.\Bar{ds} = \int_{c} \frac{d\Bar{u}}{dt}.\Bar{ds} + \int_{c} {\Bar{u}}.\frac{d\Bar{(ds)}}{dt} \tag{11}
    \end{gather*}
    In the above integral, $c$ is the closed curve enclosing the control mass, $CM$. The second integral on the right side vanishes as $\int_{c} {\Bar{u}}.({d\Bar{(ds)}}/{dt}) = \int_{c} {\Bar{u}}.{d\Bar{u}}=0$. 
    The first integral of the above equation can be estimated by substituting for ${d\Bar{u}}/{dt}$ from the momentum conservation equation as below,
    \begin{gather*}
        \frac{d\Bar{u}}{dt} = -\frac{\nabla p}{\rho} + \Bigl( \frac{\rho - \rho_a}{\rho} \Bigr) \Bar{g} + \nu \nabla^2\Bar{u} \tag{12}
    \end{gather*}
    Since the jet flame is expanding into the open environment, $\nabla p = 0$ in the above equation. Neglecting the effects of viscous dissipation (\cite{xia_vortex-dynamical_2018}), the above equation simplifies to,
    \begin{gather*}
        \frac{d\Bar{u}}{dt} = \Bigl( \frac{\rho - \rho_a}{\rho} \Bigr) \Bar{g} \tag{13}
    \end{gather*}
    Substituting for ${d\Bar{u}}/{dt}$ in Equation (11), ${d\Gamma_{CM}}/{dt}$ simplifies to,
    \begin{gather*}
         \frac{d\Gamma_{CM}}{dt} = \int_{c} \Bigl( \frac{\rho - \rho_a}{\rho} \Bigr) \Bar{g}.\Bar{ds} \tag{14}
    \end{gather*}
    The above integral vanishes along the horizontal boundaries at $x = L$ and $x = U$ since $\Bar{g}$ is orthogonal to the infinitesimal line element, $\Bar{ds}$, along these boundaries. Along the vertical boundary on the ambient side, the integrand reduces to zero since $\rho=\rho_a$. Thus, ${d\Gamma_{CM}}/{dt}$ reduces to a line integral along the boundary $F$, which simplifies to,
    \begin{gather*}
        \frac{d\Gamma_{CM}}{dt} = -g\left(\frac{\rho_a}{\rho_f}-1\right)H\left(t\right) \tag{15}
    \end{gather*}
    In the above equation, $\rho_f$ is the density of the fluid along the vertical boundary ($F$) at the flame side of the control mass. 

    Substituting for $({d\Gamma}/{dt})_{L(t)}$ and ${d\Gamma_{CM}}/{dt}$ from Equations (10) and (15) into Equation (7), we get,
    \begin{gather*}
       \frac{d\Gamma_{CV}}{dt} = -g \rho_a \left(\frac{1}{\rho_f}-\frac{1}{\rho_a}\right)H\left(t\right) - \left(\frac{u^2}{2}\right)_{x=L,r=0} \tag{16}
    \end{gather*}

    When integrated over a period of $t_{sh}$, the above equation yields the critical circulation limit at which the toroidal vortices shed. However, it requires an evaluation of $H(t)$. Schlieren flow visualization images were used to track the variation of $H(t)$ over time, and the trends observed are plotted in Figure \ref{fig:Ht_variation}(a,b). The plots are presented on the log scale to illustrate the relationship between $H(t)$ and time. The slopes of the curve (between $H(t)$ and $t$) are found to tend towards a value close to 1, suggesting a near-linear relationship between $H(t)$ and time. This observation was found to be true under nominal conditions and under the influence of the induced flow. Furthermore, $H(t)$ was also found to exhibit an increasing trend with $v_{j}$ (Figure \ref{fig:Ht_variation}(a)) under nominal conditions and was found to increase with $v_{j}$ and $v_{in}$ under the influence of the induced flow (Figure \ref{fig:Ht_variation}(b)). 

    \begin{figure}
        \centering
        \includegraphics[width=0.9\linewidth]{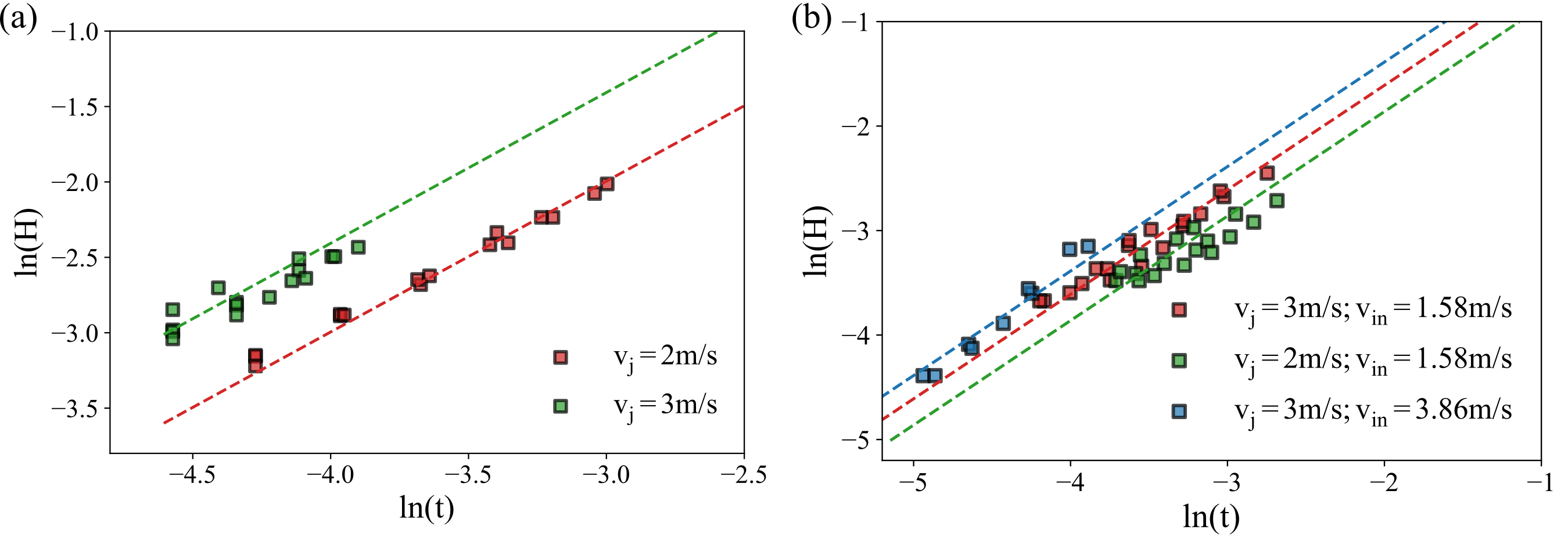}
        \caption{Panels (a,b) depict the temporal variation of $H(t)$ under nominal conditions and under the influence of the induced flow, respectively. The plots are presented on a logarithmic scale, with their slopes close to $1$, suggesting a linear relationship between $H(t)$ and time.}
        \label{fig:Ht_variation}
    \end{figure}

    Therefore, we hypothesise a correlation of the following form for $H(t)$,
    \begin{align*}
        H(t) = A t v_{j}^B \! &\text{   : Under Nominal Conditions} \\
        H(t) = F t v_{j}^G v_{in}^H \! &\text{  : Under the influence of $v_{in}$} \tag{17}
    \end{align*}
    In the above correlation, $A$, $B$, $F$, $G$, and $H$ are constants and are evaluated using the experimental data, as illustrated further in this section. 

    \begin{figure}
        \centering
        \includegraphics[width=0.9\linewidth]{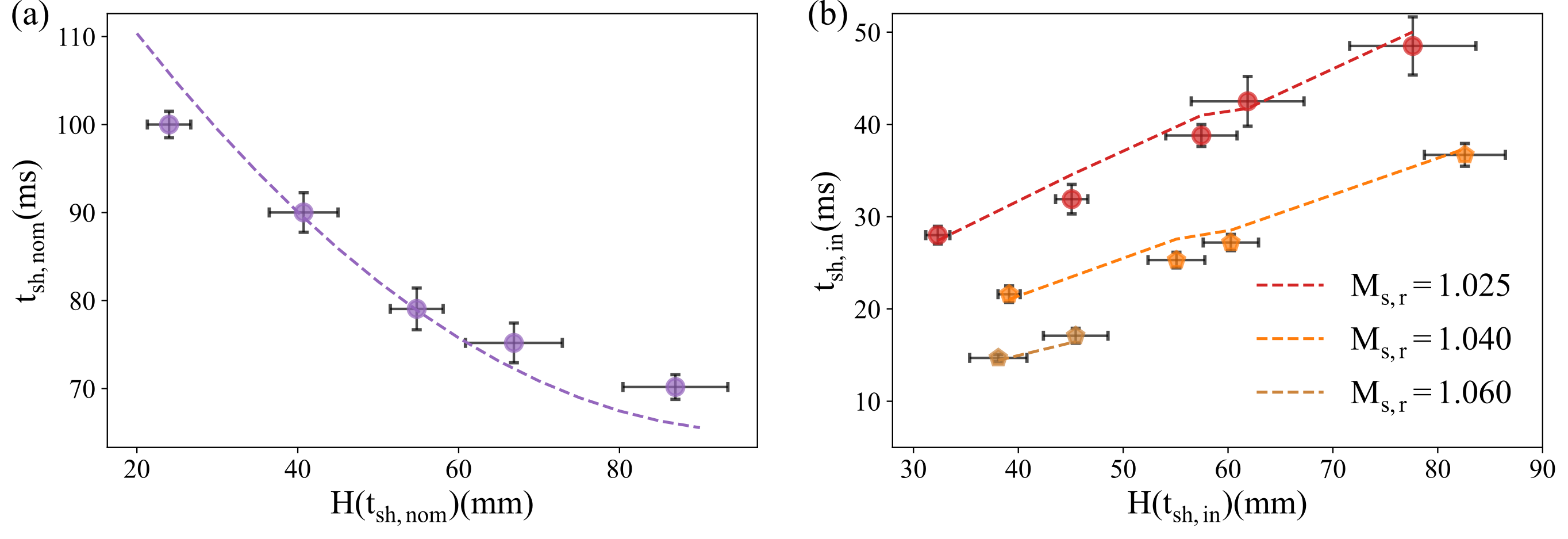}
        \caption{(a,b) The plots illustrate the variation of the shedding timescale ($t_{sh}$) against $H(t_{sh})$ in quiescent conditions and following the interaction with the induced flow, respectively. The dotted lines in these panels correspond to Equations (18) and (19), respectively.}
        \label{fig:Shedding_time_height}
    \end{figure}
    
    The plot illustrating variation of $H(t_{sh,nom})$ ($H(t)$ at the instant of shedding, under nominal conditions) against $t_{sh,nom}$ (timescale associated with shedding under nominal conditions) across the parametric space of $Re$ is presented in Figure \ref{fig:Shedding_time_height} (a). It should be noted that in a time period of $t_{sh,nom}$, the vortex convects downstream by a distance of $H(t_{sh,nom})$. Thus, the functional variation depicted in the plot should correspond to the correlation proposed in Equation (17) at the instant of vortex shedding ($t_{sh,nom}$). 
    
    Classically, the correlation between $H(t_{sh})$ and $t_{sh}$ for flickering flames is specified in terms of Froude number ($Fr=v^2/(gd)$), which essentially captures the relative dominance of the jet inertia against the buoyancy forces (\cite{xia_vortex-dynamical_2018}). Introducing a similar non-dimensional number and curve fitting the plot depicted in Figure \ref{fig:Shedding_time_height} (a) yields the following correlation:
    \begin{gather*}
        H(t_{sh,nom}) = \frac{v_j t_{sh,nom}}{k_1} \Bigl(\frac{\sqrt{gd}}{v_j}\Bigr)^a \tag{18}
    \end{gather*}
    In the above correlation, $k_1=9.85$ and $a=-0.45$. Comparing Equation (18) against the correlation proposed in Equation (17) at $t=t_{sh}$ shows that $A=((gd)^{(a/2)})/k_1$ and $B=1-a$. Substituting for $A$ and $B$ in Equation (17) thus yields a functional form for $H(t)$.

    A similar process can be used to estimate the constants $F$, $G$ and $H$ to formulate $H(t)$ under the influence of the induced flow. The plot in Figure \ref{fig:Shedding_time_height} (b) depicting the variation of $H(t_{sh,in})$ ($H(t)$ at the instant of shedding, under the influence of the induced flow) against $t_{sh,in}$ (timescale associated with shedding following the interaction with the induced flow) reveals that alongside $v_{j}$, $v_{in}$ has a significant role to play in the functional relationship between $H(t_{sh,in})$ and $t_{sh,in}$. Employing a similar curve-fitting routine as described previously yields the correlation presented in Equation (19) that captures the variation of $H(t_{sh,in})$ against $t_{sh,in}$, contingent on the operating $v_{j}$ and $v_{in}$. 
    
    \begin{gather*}
        H(t_{sh,in}) = \frac{v_j t_{sh,in}}{k_2} \Bigl(\frac{\sqrt{gd}}{v_j}\Bigr)^c \Bigl(\frac{\sqrt{gd}}{v_{in}}\Bigr)^e \tag{19}
    \end{gather*}
    In the above equation, $k_2=0.25$, $c=0.75$ and $e=-0.25$. Comparing the above correlation with that depicted in Equation (17) at the instant of shedding ($t=t_{sh,in}$) gives $F=((gd)^{(c+e)/2})/k_2$, $G=1-c$ and $H=-e$.

    We have thus obtained the following functional forms for $H(t)$ under nominal conditions and under the influence of the induced flow.
    \begin{align*}
        H(t) = \frac{v_j t}{k_1} \Bigl(\frac{\sqrt{gd}}{v_j}\Bigr)^a \! &\text{   : Nominal} \\
        H(t) = \frac{v_j t}{k_2} \Bigl(\frac{\sqrt{gd}}{v_j}\Bigr)^c \Bigl(\frac{\sqrt{gd}}{v_{in}}\Bigr)^e \! &\text{  : $v_{in}>0$} \tag{20}
    \end{align*}

    The above-mentioned correlation for $H(t)$ can now be used in Equation (16) to obtain an estimate of the critical circulation limit. Under nominal conditions, $u_{x=L,r=0}$ can be scaled with $v_{j}$ (the velocity scaling was earlier employed by \cite{didden_formation_1979} and \cite{xia_vortex-dynamical_2018}). Thus, ${d\Gamma_{CV,nom}}/{dt}$ (${d\Gamma_{CV}}/{dt}$ under nominal conditions) reduces to, 
    \begin{gather*}
       \frac{d\Gamma_{CV,nom}}{dt} = -g \rho_a \left(\frac{1}{\rho_f}-\frac{1}{\rho_a}\right) \frac{v_j t}{k_1} \Bigl(\frac{\sqrt{gd}}{v_j}\Bigr)^a - \frac{v_{j}^{2}}{2} \tag{21}
    \end{gather*}
    Integrating the above equation for a time period of $t_{sh,nom}$, we get,
    \begin{gather*}
        \Gamma_{cri,nom} = \int_{0}^{t_{sh,nom}} \frac{d\Gamma_{CV,nom}}{dt} = \\
        -\Biggl[\frac{\rho_ag}{2}\left(\frac{1}{\rho_f}-\frac{1}{\rho_a}\right) k_1 \left((gd)^{-a/2}\right) \left(v_{j}^{a-1}\right) H_{sh,nom}^2 \\
        + \frac{k_1}{2} \left((gd)^{-a/2}\right) \left(v_{j}^{a+1}\right) H_{sh,nom}\Biggr] \tag{22}
    \end{gather*}

    A similar expression can be obtained by considering vortex shedding under the influence of the induced flow. Substituting the correlation for $H(t)$ from Equation (20) (for $v_{in}>0$) into Equation (16), and using $(v_{j}+v_{ind})$ as the velocity scale for $u_{x=L,r=0}$, we get,
    \begin{gather*}
       \frac{d\Gamma_{CV,in}}{dt} = -g \rho_a \left(\frac{1}{\rho_f}-\frac{1}{\rho_a}\right) \frac{v_j t}{k_2} \Bigl(\frac{\sqrt{gd}}{v_j}\Bigr)^c \Bigl(\frac{\sqrt{gd}}{v_j}\Bigr)^e - \frac{(v_{j}+v_{in})^{2}}{2} \tag{23}
    \end{gather*}
    It is important to note that in order to scale $u_{x=L,r=0}$ as $(v_{j}+v_{ind})$, the lower boundary ($L$) must be positioned at an axial distance of $x_{L}$ (Figure \ref{fig:Shedding_schematic}(a)), which takes into account the effect of recirculation of the induced flow as it traverses across the nozzle exit. For the operating conditions corresponding to the current experiments (characterized by the Reynolds number based on $v_{in}$ and $d$), the length of the recirculation zone ($x_{L}$) is approximately equal to $d$ (\cite{taneda_experimental_1956}). Consequently, it is safe to assume that the induced flow imposes a velocity field that scales with $v_{in}$ at $r=0$, thereby making the effective velocity to scale with $(v_{j}+v_{ind})$. It is worth mentioning that the spatial positioning of $L$ at $x=x_{L}$ would not alter the formulation of critical circulation under nominal conditions (Equation (22)), as the scale of $u_{x=L,r=0}$ would remain $v_{j}$ for $x_{L} \sim d$. It should also be noted that $x_{L} << H(t_{sh})$ under nominal conditions and under the influence of the induced flow.
    
    Integrating Equation (23) over a time period of $t_{sh,in}$, we can thus obtain an estimate for critical circulation for shedding following the interaction with the induced flow as,
    \begin{gather*}
       \Gamma_{cri,in} = \int_{0}^{t_{sh,in}} \frac{d\Gamma_{CV,in}}{dt} = \\
       -\Biggl[\frac{\rho_ag}{2}\left(\frac{1}{\rho_f}-\frac{1}{\rho_a}\right) k_2 \left((gd)^{-(c+e)/2}\right) \left(v_{j}^{c-1}\right) \left(v_{in}^{e}\right) H_{sh,in}^2 \\
       + \frac{k_2}{2} ((v_{j}+v_{in})^2) \left((gd)^{-(c+e)/2}\right) \left(v_{j}^{c-1}\right) \left(v_{in}^{e}\right) H_{sh,in} \Biggr] \tag{24}
    \end{gather*}

    We can now equate Equations (22) and (24) following the hypothesis of \cite{pandey_dynamic_2021} and \cite{thirumalaikumaran_insight_2022} that the critical circulation limit at which the toroidal vortices shed in flickering flames remains constant irrespective of the changes imposed over the flame by an external flow. We can thus obtain an expression to theoretically evaluate the shedding heights in the presence of the induced flow ($h_{sh,in,th}$).

    \begin{gather*}
       \Biggl[\frac{\rho_ag}{2}\left(\frac{1}{\rho_f}-\frac{1}{\rho_a}\right) k_1 \left((gd)^{-a/2}\right) \left(v_{j}^{a-1}\right) H_{sh,nom}^2 \\
        + \frac{k_1}{2} \left((gd)^{-a/2}\right) \left(v_{j}^{a+1}\right) H_{sh,nom}\Biggr] = \\
       \Biggl[\frac{\rho_ag}{2}\left(\frac{1}{\rho_f}-\frac{1}{\rho_a}\right) k_2 \left((gd)^{-(c+e)/2}\right) \left(v_{j}^{c-1}\right) \left(v_{in}^{e}\right) H_{sh,in,th}^2 \\
       + \frac{k_2}{2} ((v_{j}+v_{in})^2) \left((gd)^{-(c+e)/2}\right) \left(v_{j}^{c-1}\right) \left(v_{in}^{e}\right) H_{sh,in,th} \Biggr] \tag{25}
    \end{gather*}

    Solving the above equation for $H_{sh,in,th}$ yields a theoretical estimate for the shedding height under the influence of the induced flow. 
    Figure \ref{fig:Shedding_height_trends}(a) compares $H_{sh,in,th}$ (scatter plot with solid lines) and experimentally observed values of $H_{sh,in}$ (scatter plot with dashed lines and error bars) across the reattachment regimes. The theoretical model is found to overestimate the shedding heights with a maximum error margin of $35\%$ (observed at the lowest values of $v_{j}$ and $v_{in}$ ). The discrepancies between experimental results and theoretical predictions are found to diminish with increasing fuel jet and induced flow velocities. The overestimation in the theoretical prediction likely stems from the underestimation of the induced velocity scale, as determined in this study based on the flame base lift-off rate at $t_{0}$. Additionally, the current theoretical model neglects the effect of the pressure gradient imposed by the induced flow. Accurate estimation of these parameters was not feasible due to the limitations of our experimental facility.

    Thus, Equation (25), alongside Equation (19), can be used to theoretically predict the shedding length and timescales, contingent on the operating conditions of $v_{j}$ and $v_{in}$. The developed model is subsequently employed in section \ref{subsec:Extinction} to estimate $H(t_{sh,in,th})$ and $t_{sh,th}$ in the extinction regimes to account for a few experimental observations. 

    \begin{figure}
        \centering
        \includegraphics[width=0.9\linewidth]{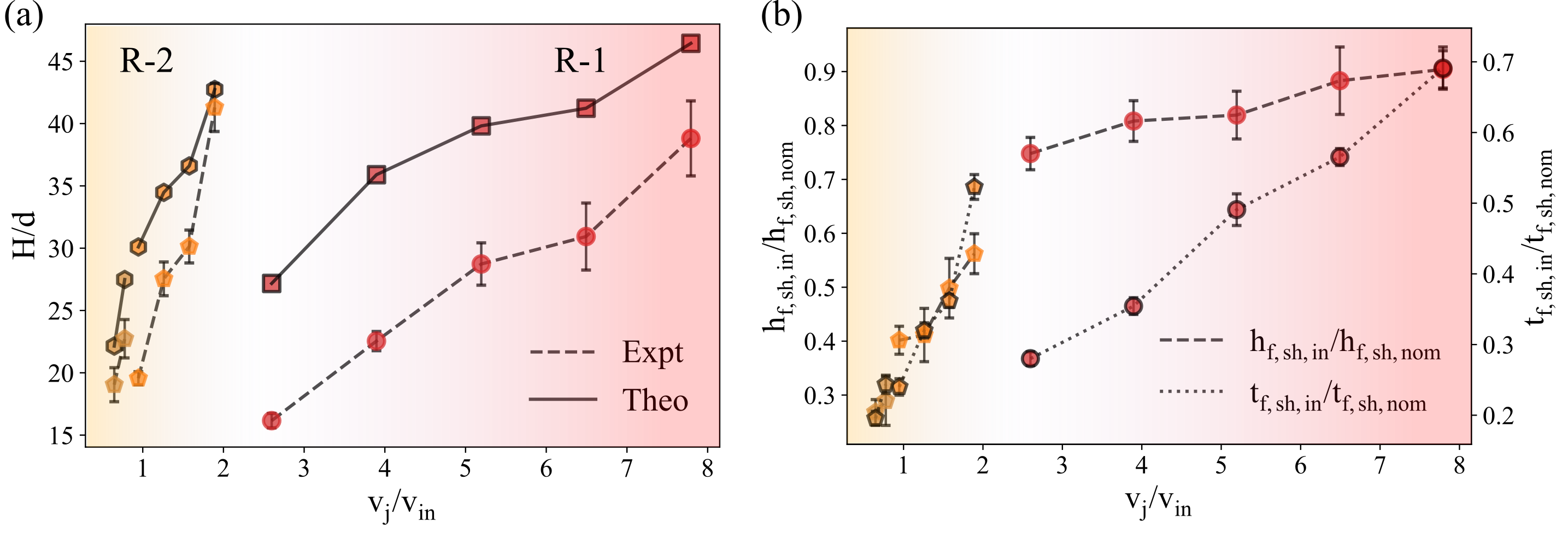}
        \caption{(a) Comparison of the observed experimental trends (plotted in dotted lines) in $H(t_{sh,in})$ against corresponding theoretical estimates (represented by solid lines). (b) The plot compares the observed shedding heights and timescales when the jet flame interacts with the induced flow to those observed under quiescent conditions.}
        \label{fig:Shedding_height_trends}
    \end{figure}

    Figure \ref{fig:Shedding_height_trends}(b) compares the height of the flame at the instant of shedding in the presence of the induced flow ($h_{f,sh,in}$) against those observed under nominal conditions ($h_{f,sh,nom}$). It is to be noted that $h_{f,sh,in}$ is the distance between the flame tip and the flame base and is different from $H(t_{sh,in})$ (height of the control volume enclosing the toroidal vortex as measured from a fixed lower boundary, $L$) for flames responding to induced flow since the flame base lifts off as it interacts with the induced flow. 
    The plot (Figure \ref{fig:Shedding_height_trends}(b)) clearly depicts that the decrease in $h_f$ is more pronounced in the reattachment-2 regime compared to the reattachment-1 regime (plotted in red). This difference illustrates the reason for the observed phenomena of flame necking and pinch-off in the reattachment-2 regime. As the toroidal vortex reaches its critical circulation limits at heights significantly lower than $h_{f,sh,nom}$ in the reattachment-2 regime, it tends to pinch off a portion of the flame tip as it sheds. However, in the reattachment-1 regime, the flame height at the instant of shedding ($h_f{sh,in}$) is comparable to that observed under nominal conditions ($h_{f,sh,nom}$), causing the flame tip to experience stretching or deformation rather than a pinch-off event. The ratio of the shedding timescales, $t_{sh,in}/t_{sh,nom}$, also exhibit a similar trend as $h_f(t_{sh,in})/h_f(t_{sh,nom})$, wherein the drop in the shedding timescale is significantly higher in the reattachment-2 regime in comparison with the reattachment-1 regime.

    \subsection{Extinction Response Regimes}\label{subsec:Extinction} \addvspace{10pt}
   
    The flame response to the induced flow in the Extinction Type-1 and Type-2 regimes is illustrated in Figure \ref{fig:E1_E2_Quant}(a,b), respectively. Similar to the reattachment regimes, the distinction between Type-1 and Type-2 extinction regimes is primarily based on the flame tip response. In the Type-2 sub-regime, a distinct neck formation event is observed at the flame tip, which is absent in the Type-1 sub-regime. However, unlike the reattachment-2 regime, flame shedding was not consistently observed (statistical occurrence during experiments) in the extinction-2 regime, although neck formation was present. This is due to an additional competing timescale associated with flame extinction in the extinction regimes.
 
    In both extinction regimes, the flame base continues to lift off upon encountering the induced flow without attaining a maximum (Figure \ref{fig:E1_E2_Quant}(a,b)). As illustrated in Figure \ref{fig:Global_All_Flames}(d,e), this leads to an eventual interaction between the flame base and the flame tip in the extinction Type-1 regime and causes flame base-flame neck interaction in the type-2 regime. We hypothesise that the interaction causes flame strain rates to exceed critical strain rate limits and is responsible for flame extinction. Validating the hypothesis, however, requires velocity data on the evolving flow field, which is beyond the scope of the present work. It is also observed that the OH* chemiluminescence signal of the flame (that scales with the flame’s heat release rate) monotonically starts to drop beyond the interaction phase (shaded region in Figure \ref{fig:E1_E2_Quant}(a,b)).
    
    In Figure \ref{fig:Lift_off_shed_ext}(a), a comparison between the flame base lift-off height (normalised with the $h_{f,sh,nom}$) is presented across the flame response regimes from reattachment to extinction. The plots correspond to the maxima of the flame base lift-off height in the reattachment regimes ($h_{b,lft}$) and the flame base lift-off height at which flame extinction is observed ($h_{b,lft,ext}$) in the extinction regimes. It is evident from the figure that the flame base lift-off height is higher in the extinction regimes than in the reattachment regimes. While the flame base lifts off to heights comparable to the nominal flame height in the extinction-1 regime, causing an interaction between the flame base and flame tip (Figure \ref{fig:Global_All_Flames}(d)), $h_{b,lft,ext}/h_f(t_{sh,nom})$ varies between 0.55 and 0.8 in the extinction-2 regime wherein the flame base interacts with the flame neck location (Figure \ref{fig:Global_All_Flames}(e)) prior to extinction. 

    \begin{figure}
        \centering
        \includegraphics[width=0.9\linewidth]{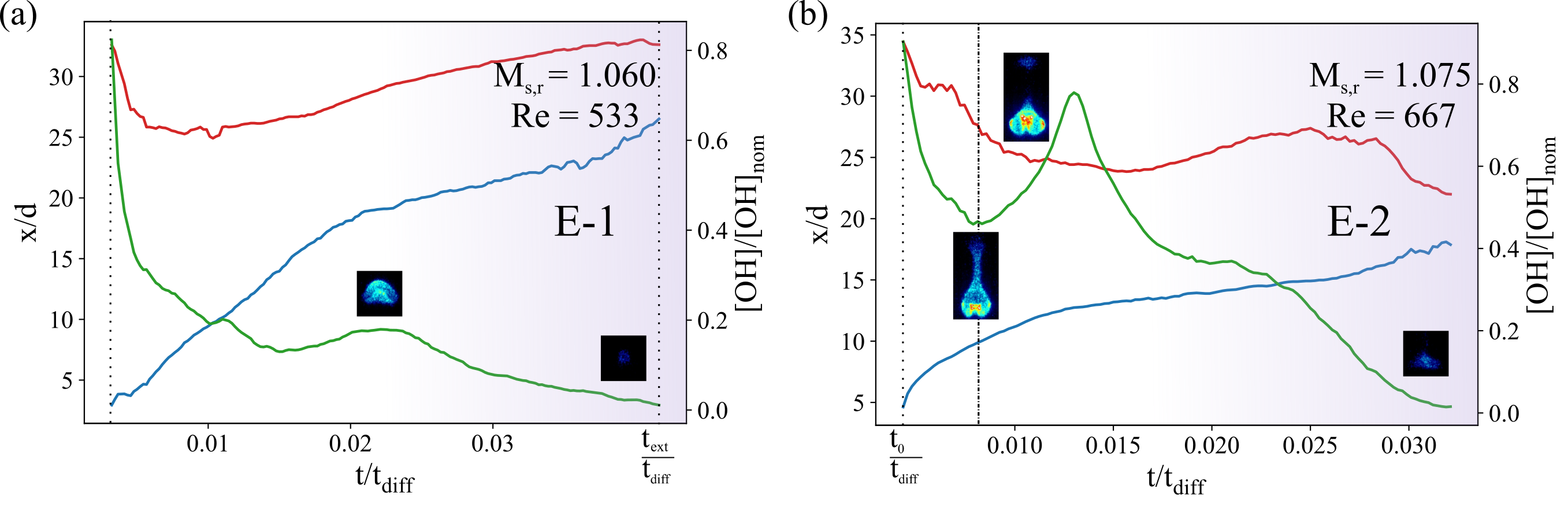}
        \caption{The figure depicts the temporal response of the flame base (blue) and the flame tip (red) alongside the OH* chemiluminescence signal of the jet flame as it interacts with the blast wave and the subsequent induced flow within the extinction regime. Panels (a,b) correspond to the extinction sub-regimes, Type-1 and Type-2, respectively. The shaded region in the plots represents the timescale range over which a monotonic decrease in the OH chemiluminescence signal is observed following the interaction of the flame base with the flame tip in the Type-1 sub-regime and with the flame neck location in the Type-2 sub-regime.}
        \label{fig:E1_E2_Quant}
    \end{figure}

    \begin{figure}
        \centering
        \includegraphics[width=0.9\linewidth]{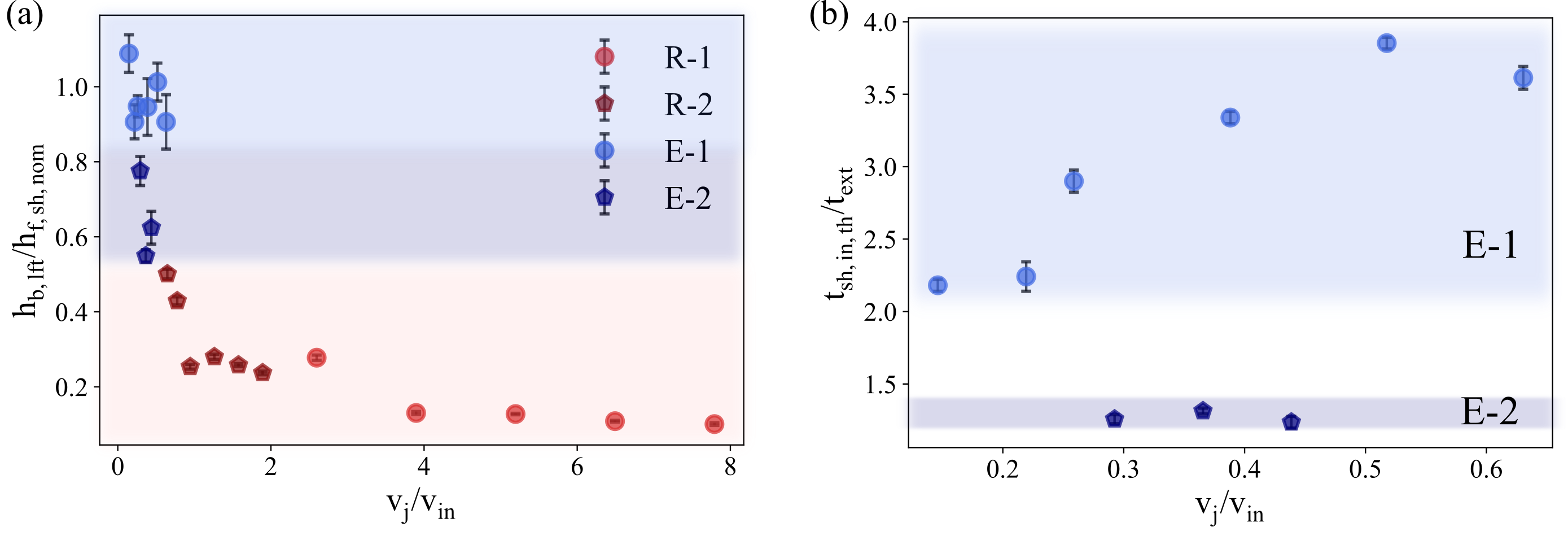}
        \caption{(a) The plot compares the observed maximum flame base lift-off height against the nominal shedding height at the corresponding fuel jet Reynolds number. (b) The plot compares the theoretically estimated shedding timescales in the extinction regimes against the timescale associated with flame extinction.}
        \label{fig:Lift_off_shed_ext}
    \end{figure}
  
    Flame shedding is observed in the extinction-2 regime only when the time required for the toroidal vortex to reach its critical circulation limit ($t_{sh,in}$) is lower than the extinction time scale ($t_{ext}$). However, if $t_{sh,in}$ is greater than $t_{ext}$, flame pinch-off is not observed, although a neck formation might still be present. Extending the formulation developed in Section \ref{subsec:Reattachment}, we can estimate $H(t_{sh,in,th})$ and $t_{sh,in,th}$ for the extinction regimes. Figure \ref{fig:Lift_off_shed_ext}(b) compares the estimated shedding timescales against the experimentally observed extinction time scale ($t_{ext}$). It is to be noted that the theoretical model developed in Section \ref{subsec:Reattachment} tends to overestimate $H(t_{sh,in,th})$, as evident in Figure \ref{fig:Shedding_height_trends}(a). Correspondingly, $t_{sh,in,th}$ is also expected to be overestimated since there exists a linear correlation between $H(t_{sh,in,th})$ and $t_{sh,in,th}$ (Equation (19)). Thus, the values of $t_{sh,in,th}$ in Figure \ref{fig:Lift_off_shed_ext}(b) are expected to be overestimated. The ratio between $t_{sh,in,th}$ and $t_{ext}$ lies between 1.2 and 1.3 in the extinction-2 regime and between 1.9 and 3.8 in the extinction-1 regime. $t_{sh,in,th}$ is thus comparable to $t_{ext}$ in the extinction-2 regime and explains the statistical observation of flame shedding in the experimental trials. However, $t_{ext}$ is considerably smaller than $t_{sh,in,th}$ in the extinction-1 regime, and therefore, the flame undergoes extinction without exhibiting a flame shedding/pinch-off event.

    \subsection{Regime Map}\label{subsec:Regime_Map} \addvspace{10pt}

    Figure \ref{fig:Regime_Map} illustrates the parametric space of fuel jet Reynolds numbers and incident Blast wave Mach numbers ($M_{s,r}$) over which different flame response regimes were observed. Reducing the fuel jet Reynolds number and increasing the blast strength were found to make the flame vulnerable to extinction. Reattachment Type-1 response was restricted to the lowest Blast strength ($M_{s,r}$=1.025) across the entire range of Re. The reattachment-2 regime was noted for $Re\geq400$ and $Re\geq667$ at blastwave Mach numbers of 1.040 and 1.060, respectively. Extinction-1 regime was dominant in shorter flames and was observed till the Reynolds number of 267, 534 and 400 at blast wave Mach numbers ($M_{s,r}$) of 1.040, 1.060 and 1.075, respectively. Extinction-2 sub-regime was observed at the highest blast strength ($M_{s,r}$=1.075) for $Re\geq534$.

    Type-2 response was typically observed in longer flames (higher $Re$) and higher induced velocities ($M_{s,r} \geq 1.040$), wherein the toroidal vortices have enough convective lengthscale and circulation build-up rate to roll up and reach critical circulation limits at heights lower than the flame heights. 

    \begin{figure}
        \centering
        \includegraphics[width=0.6\linewidth]{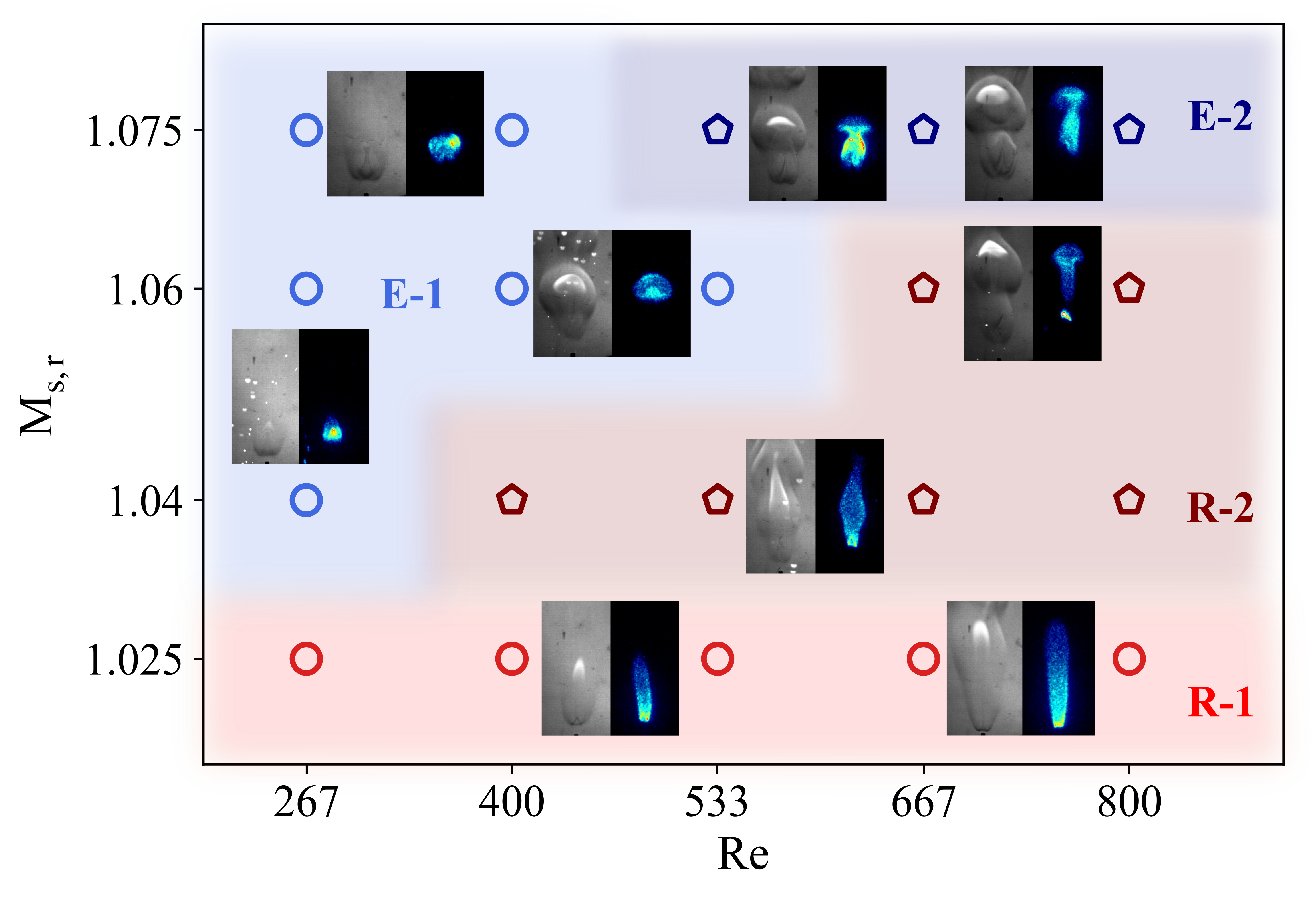}
        \caption{Regime Map illustrating the range of diverse flame response behaviours to the blast wave and the ensuing induced flow, spanning the parametric space of fuel jet Reynolds number and the incident blast wave Mach number.}
        \label{fig:Regime_Map}
    \end{figure}

\section{Conclusion}\label{sec:conclusion}

The study investigates the interaction dynamics of non-premixed jet flames with blast waves. As a blast wave sweeps across an attached non-premixed jet flame along the jet axis, it imposes a characteristic flow field marked by a sharp discontinuity followed by a decaying profile and a delayed second spike. The second spike corresponds to the induced flow that follows the blast front. While the flame responds to the shock front with a jittery motion, it tends to lift off from the nozzle rim following the interaction with the induced flow. The lifted flame is found to exhibit two global response patterns. One, wherein the flame reattaches back at the nozzle rim in a time period of $t_{b,ra}$ after lifting off to a characteristic lift-off height of $h_{b,lft}$, and the other, wherein the flame extinguishes following the interaction of the flame base with the flame tip or the flame neck. Furthermore, the dynamics of flame tip flickering were altered following the interaction with the induced flow. Within a characteristic parametric range of fuel jet Reynolds numbers ($Re$) and incident blast wave Mach numbers ($M_{s,r}$), the flame tip exhibited a neck formation. As a result, the re-attachment and extinction regimes were further subdivided into Type-1 and Type-2 sub-regimes, where neck formation at the flame tip was absent in Type-1 and present in Type-2.

A formulation extending the vorticity transport equation was used to develop a theoretical model to estimate the change in flickering timescales and length scales of the jet flame following its interaction with the induced flow. The model posits that tip flickering results from the periodic shedding of toroidal vortices, which develop along the shear boundary between the hot product gases and the ambient air due to gravity-induced shearing. Under the influence of the induced flow, the circulation build-up rate of these vortices is enhanced, and they tend to reach their critical circulation limits (at which they shed) at heights lower than their nominal shedding heights. Flame pinch-off or shedding was observed only when the shedding height dropped below the height of the flame. This causes a portion of the flame tip to pinch off along with the vortex as it sheds. 

The developed theoretical model was extended into the parametric space of the extinction regime. A criterion based on the ratio of the timescale associated with circulation build-up and the extinction timescale was formulated to differentiate between the Type-1 and Type-2 sub-regimes within the extinction regime. 
 
\section*{Declaration of competing interest} \addvspace{10pt}

The authors have no competing interest to disclose

\section*{Acknowledgments} \addvspace{10pt}

The authors are thankful to SERB - CRG (CRG/2020/000055) for financial support. S.B. acknowledges funding through the Pratt and Whitney Chair Professorship. A.A. acknowledges funding received through the PMRF Fellowship scheme.

\bibliographystyle{jfm}
\bibliography{jfm-instructions}

\end{document}